\documentclass[11pt,a4paper]{article}
\pdfoutput=1
\usepackage{jheppub}
\usepackage{amsmath,bm,braket}
\usepackage{graphicx}
\usepackage{caption}
\usepackage{subcaption}

\newcommand{\nb}{\bar{n}}
\newcommand{\msbar}{$\overline{\text{MS}}$}
\newcommand{\Acal}{\mathcal{A}}
\newcommand{\Bcal}{\mathcal{B}}
\newcommand{\Ical}{\mathcal{I}}
\newcommand{\sO}{\mathcal{O}}
\newcommand{\as}{\alpha_s}
\newcommand{\asFPi}{\frac{\as}{4\pi}}
\newcommand{\lp}{L_\perp}
\newcommand{\xp}{x_\perp}
\newcommand{\ib}{\bar{\imath}}
\newcommand{\tcdot}{\!\cdot\!}
\newcommand{\eqn}[1]{eq.~\eqref{#1}}

\preprint{DESY 14-037, ZU-TH 11/14}

\title{Calculation of the transverse parton distribution functions at next-to-next-to-leading order}

\author[b]{Thomas Gehrmann,}
\author[b,c]{Thomas L\"ubbert}
\author[a,d,e]{and Li Lin Yang}

\affiliation[a]{School of Physics and State Key Laboratory of Nuclear Physics and Technology, Peking University, Beijing 100871, China}
\affiliation[b]{Department of Physics, University of Z\"urich, 8057 Z\"urich, Switzerland}
\affiliation[c]{II.\ Institute for Theoretical Physics, University of Hamburg, 22761 Hamburg, Germany}
\affiliation[d]{Collaborative Innovation Center of Quantum Matter, Beijing, China}
\affiliation[e]{Center for High Energy Physics, Peking University, Beijing 100871, China}

\emailAdd{thomas.gehrmann@uzh.ch}
\emailAdd{thomas.luebbert@desy.de}
\emailAdd{yanglilin@pku.edu.cn}

\abstract{We describe the perturbative calculation of the transverse parton distribution functions in all partonic channels up to next-to-next-to-leading order based on a gauge invariant operator definition.
We demonstrate the cancellation of light-cone divergences and show that universal process-independent transverse parton distribution functions can be obtained through a refactorization.
Our results serve as the first explicit higher-order calculation of these functions starting from first principles, and can be used to perform next-to-next-to-next-to-leading logarithmic $q_T$ resummation for a large class of processes at hadron colliders.}

\begin{document}

\maketitle
\newpage

\section{Introduction}

Parton distribution functions (PDFs) are fundamental properties of hadrons. They describe the distributions of quarks and gluons inside hadrons. Their usefulness in collider phenomenology resides in the factorization theorems, in which short range interactions are separated from long range effects. The short range interactions result in perturbatively calculable hard scattering kernels or hard functions, while the long range effects are encoded in non-perturbative or semi-perturbative objects such as soft functions, parton distribution functions and fragmentation functions. From these, it is possible to make predictions for collider observables based on perturbative calculations and experimental determination of the non-perturbative functions. This procedure has proven highly successful in the last thirty years.

For most applications, the relevant factorization theorems are ``collinear factorization'' developed in \cite{Collins:1985ue, Bodwin:1984hc, Collins:1989gx}. The corresponding non-perturbative functions are so-called ``collinear PDFs'' or ``integrated PDFs'', in which only the partonic momentum component along the direction of the colliding hadron is kept, while all other components are integrated over.
The collinear PDFs have gauge-invariant definitions in terms of matrix elements of non-local bilinear operators \cite{Collins:1981uw, Collins:2011zzd}. While the collinear PDFs are non-perturbative functions, their scale-dependence can be calculated perturbatively in terms of the DGLAP splitting kernels.
These calculations have been performed up to 3 loops \cite{Altarelli:1977zs, Dokshitzer:1977sg, Gribov:1972ri, Curci:1980uw, Furmanski:1980cm, Moch:2004pa, Vogt:2004mw}.

In many circumstances, in addition to the partonic momentum component collinear to the hadron, other momentum components and possibly also the polarization of the parton can be important.
Therefore, it is necessary to generalize the collinear factorization to incorporate these degrees of freedom, which leads to generalized factorization theorems in terms of generalized PDFs and fragmentation functions.
Popular examples are virtuality dependent factorization \cite{Stewart:2010qs,Gaunt:2014xga},
transverse momentum dependent (TMD) factorization \cite{Collins:1981uk, Collins:1981uw, Collins:1984kg, Catani:2000vq, Ji:2004wu, Ji:2004xq, Bozzi:2005wk, Becher:2010tm, Becher:2011dz, Chiu:2011qc, GarciaEchevarria:2011rb, Collins:2011zzd, Chiu:2012ir, Becher:2012yn}, 
as well as virtuality and transverse momentum dependent factorization \cite{Collins:2007ph,Mantry:2009qz,Jain:2011iu}.

In this work, we will consider TMD factorization, where transverse parton distribution functions (TPDFs) and transverse fragmentation functions are introduced.
Historically, TMD factorization frameworks were developed in three different kinds of kinematics: $e^+e^-$ to hadrons, semi-inclusive deep-inelastic-scattering (SIDIS), and Drell-Yan type processes.
In this work, we will mainly be concerned with hadron collider physics, and will therefore discuss TMD factorization and TPDFs for (unpolarized) Drell-Yan type processes in detail.

Consider the production of a vector boson in hadron-hadron collisions with its invariant mass $Q$ and transverse momentum $q_T$ observed. If $Q \sim q_T \gg \Lambda_{\text{QCD}}$, one expects that collinear factorization is valid and the differential cross section can be factorized as (schematically)
\begin{align}
  \label{eq:colfac}
  \frac{d^2\sigma}{dQdq_T} \sim \phi_{i/N_1}(x_1,\mu) \otimes \phi_{j/N_2}(x_2,\mu) \otimes C_{ij}(z,Q,q_T,\mu) \, .
\end{align}
In the above, $\phi_{i/N}(x,\mu)$ are the collinear PDFs with the longitudinal momentum fraction $x$. They are non-perturbative functions describing physics at the hadronic scale $\Lambda_{\text{QCD}}$. $C_{ij}(z,Q,q_T,\mu)$ are hard scattering kernels describing physics at the hard scale $Q \sim q_T$. The symbol $\otimes$ denotes convolution.

Consider now another phenomenologically important region $Q \gg q_T, \Lambda_{\text{QCD}}$. In this region, even if $q_T$ is in the perturbative domain, use of the collinear factorization formula (\ref{eq:colfac}) will lead to problems with the perturbation series due to the appearance of large logarithms of $Q/q_T$.
Therefore, one would like to factorize the two scales and resum the large logarithms to all orders in perturbation theory. Ideally, one might expect a factorization formula similar to \eqn{eq:colfac}:
\begin{align}
  \frac{d^2\sigma}{dQdq_T} \sim \tilde{\Bcal}_{i/N_1}(x_1,k_{1T},\mu) \otimes \tilde{\Bcal}_{j/N_2}(x_2,k_{2T},\mu) \otimes H_{ij}(z,Q,\mu) \, ,
\end{align}
where the hard functions $H_{ij}(z,Q,\mu)$ describe physics at the hard scale $Q$, and the TPDFs $\tilde{\Bcal}_{i/N}(x,k_T,\mu)$ describe physics at the low scales $q_T$ and $\Lambda_{\text{QCD}}$. However, things are not so simple.
It turned out that the $\tilde{\Bcal}$ functions necessarily depend on the hard scale $Q$ through some unphysical parameter, denoted here by $\xi$. Moreover, depending on the regulator used and the process under consideration, an additional soft function may appear. Therefore, the correct formula looks like
\begin{align}
  \label{eq:fac1}
  \frac{d^2\sigma}{dQ dq_T} \sim \tilde{\Bcal}_{i/N_1}(x_1,k_{1T},\mu;\xi_1) \otimes \tilde{\Bcal}_{j/N_2}(x_2,k_{2T},\mu;\xi_2) \otimes \tilde{\mathcal{S}}_{ij}(k_{T},\mu;\xi_1,\xi_2) \otimes H_{ij}(z,Q,\mu) \, .
\end{align}
Note that while the individual functions depend on the unphysical parameters $\xi_1$ and $\xi_2$, in physical cross sections these parameters are combined in a way that only the physical scale $Q$ remains. Eq.~(\ref{eq:fac1}) is not a true factorization since the $\tilde{\Bcal}$ and $\tilde{\mathcal{S}}$ functions still involve the two widely separated scales $Q$ and $q_T$.
To achieve a proper factorization of the two scales, one needs to extract the $Q$ dependence from the $\tilde{\Bcal}$ and $\tilde{\mathcal{S}}$ functions by studying their dependence on the unphysical parameters $\xi_1$ and $\xi_2$.
This procedure has various names in the literature: Collins-Soper equation in the pioneering works \cite{Collins:1981uk, Collins:1984kg, Catani:2000vq, Ji:2004wu, Ji:2004xq, Bozzi:2005wk}, rapidity renormalization group in \cite{Chiu:2011qc, Chiu:2012ir}, and refactorization in \cite{Becher:2010tm, Becher:2012yn}. After this, the $Q$ dependence can be exponentiated and one can obtain the true TPDFs.

The anomalous $Q$ dependence of the naive TPDFs $\tilde{\Bcal}$ arises as follows. Similar to the collinear PDFs, the naive TPDFs can be defined as matrix elements of non-local operators. This was given in axial gauge in \cite{Collins:1981uw} and was rendered gauge-invariant in \cite{Ji:2004wu, Ji:2004xq, Collins:2011zzd} by introducing Wilson lines.
In these works, it was pointed out that by taking the gauge-fixing vectors to the light cone, or equivalently putting the Wilson lines on the light cone, one encounters singularities not regularized in dimensional regularization in the perturbative calculations. Therefore, to perform the calculations, one has to introduce an extra regulator.
In \cite{Collins:1981uw, Ji:2004wu, Ji:2004xq}, this was achieved by taking the gauge-fixing vectors or the Wilson lines off the light cone. Other choices of regulator are possible. For example, variations of the analytic regulator were used in \cite{Chiu:2011qc, Chiu:2012ir, Becher:2010tm, Becher:2011dz}.
In \cite{GarciaEchevarria:2011rb}, finite imaginary parts in certain propagators were used to regulate the light-cone divergences. No matter which regulator is used, the anomalous $Q$ dependence inevitably arises which, in the language of soft-collinear effective theory (SCET) \cite{Bauer:2000yr,Bauer:2001yt,Beneke:2002ph}, is due to the breaking of the rescaling invariance of the Lagrangian \cite{Becher:2010tm}. 
In the Collins-Soper approach~\cite{Collins:1984kg}, the explicit appearance of those singularities is circumvented by always considering the product of two TPDFs. Very recently, this approach was worked out in full generality to describe arbitrary color-singlet final states in hadronic collisions in~\cite{Catani:2013tia}.

Depending on the relative size of the two scales $q_T$ and $\Lambda_{\text{QCD}}$, different aspects of TPDFs can be described in perturbative QCD. 
If $q_T \sim \Lambda_{\text{QCD}}$, the TPDFs are fully non-perturbative and only their scale-dependence can be calculated perturbatively. 
In the situation $q_T \gg \Lambda_{\text{QCD}}$, the TPDFs are semi-perturbative objects and one can further factorize the two scales.
In this region, the TPDFs can be expressed as convolutions of the collinear PDFs with perturbatively calculable matching coefficient functions. These coefficient functions can be obtained via two approaches.
The first is assuming the factorization formula (\ref{eq:fac1}), and extracting the coefficient functions by studying the small $q_T$ behavior of the differential cross section. This is the approach taken by \cite{Catani:2011kr, Catani:2012qa}, which is generalized to any color-singlet process 
in~\cite{Catani:2013tia}.
The second approach is starting from a gauge-invariant operator definition of the TPDFs, and straightforwardly computing the operator matrix elements. This approach is much more challenging since one directly encounters the light-cone divergences.
However, the second approach, once accomplished, serves as an explicit verification of the TMD factorization framework.
In \cite{Gehrmann:2012ze}, we derived the next-to-next-to-leading order (NNLO) coefficient function for quark-to-quark transitions. Results at this order are for example relevant for a next-to-next-to-next-to-leading logarithmic (N$^3$LL) transverse momentum resummation.
This paper extends the calculation of the NNLO coefficient functions to all partonic channels and describes technical and methodological details. 
We also present several consistency checks on our results. In particular, we reproduce the process specific $\mathcal{H}^{(2)}$ coefficients of \cite{Catani:2011kr, Catani:2012qa} for Drell-Yan process and Higgs production, as well as the order $\as^2$ contributions to the DGLAP splitting kernels.

This paper is organized as follows. In Section~\ref{sec:framework} we introduce our calculational framework. We provide the operator definitions of the TPDFs and the regularization of the light-cone singularities.
We outline the procedure of the NNLO calculations in Section~\ref{sec:pert_calc}, with some detailed expressions collected in Appendix~\ref{sec:IRR} and \ref{sec:list_IRR}. The main results are presented in Section~\ref{sec:results}, while several additional relations are collected in Appendix~\ref{sec:ad} and \ref{sec:outsourced_results}. We conclude in Section~\ref{sec:conclusions}.

\section{Framework}
\label{sec:framework}

We consider the collision of two hadrons $N_1$ and $N_2$ with momenta $p$ and $\bar{p}$ producing some color-neutral final state $F$ of momentum $q$ and additional unresolved remnants $X$
\begin{align}
  N_1(p) + N_2(\bar{p}) \rightarrow F(q) + X \, .
  \label{eq:process}
\end{align}
Along the directions of the hadrons we specify two light-like vectors $n$ and $\nb$ with $n\cdot \nb = 2$. In terms of them any 4-vector can be decomposed as 
\begin{align}
q^\mu = \frac{n^\mu}{2}\nb \cdot q + \frac{\nb^\mu}{2} n \cdot q + q_\perp^\mu
\,,
\end{align}
where $q_\perp^\mu$ is perpendicular to both $n$ and $\nb$. We define $q_T^2=-q_\perp^2$.

We consider the differential cross section for the production of the final state $F$ with respect to its squared invariant mass $q^2$, transverse momentum $q_T$, and rapidity $y$.
We are especially interested in the region where the transverse momentum is much smaller than the invariant mass $q^2 \gg q_T^2$.
For this multi-scale problem we need to achieve the factorization of disparate scales and the resummation of the corresponding logarithms.
This was done for the Drell-Yan process in the pioneering work \cite{Collins:1984kg}. 
In this paper, we will mainly follow the SCET based language in \cite{Becher:2010tm, Becher:2012yn}, in which the factorization formula for the Drell-Yan process can be written as
\begin{align}
  \label{eq:d3sigmadMdQtdY_sBsB}
  \frac{d^3\sigma}{d q^2d q_T^2 d y} &= \frac{4\pi\alpha^2}{3N_c q^2 s} \, \left| C_V(-q^2-i\epsilon) \right|^2
  \frac{1}{4\pi} \int d^2 x_\perp \, e^{-i q_\perp\cdot x_\perp} \nonumber
  \\ 
  &\hspace{1.5cm} \times \sum_{q}  e_q^2 \left[ {\cal{S}}_{q\bar{q}}(x_T^2) \, \Bcal_{q/N_1}(z_1,x_T^2) \, \bar{\Bcal}_{\bar{q}/N_2}(z_2,x_T^2) + (q\leftrightarrow \bar{q}) \right] ,
\end{align}
where the first factor corresponds to the Born level cross section, $C_V$ is the Wilson coefficient obtained from matching the quark form factor to the effective theory. Together they form the process specific hard function. The transverse position (impact parameter) $x_\perp$ is the Fourier conjugate variable to $q_\perp$, and $x_T^2=-x_\perp^2$.
The position-space soft function $\cal{S}$ is a correlator of soft Wilson lines, and $\Bcal$ and $\bar{\Bcal}$ are the two position-space TPDFs.
In terms of $x_T^2$ they depend on the transverse variable. Other functional dependences related to the regularization are implicit in the above expression and the $-i\epsilon$ prescription for the Wilson coefficient defining the sign of its imaginary part will be suppressed from now on.
The factorization theorem holds up to power corrections in $q_T^2/q^2$.

\subsection{Definition of transverse  PDFs}

The bare quark TPDF collinear to the $n$ direction is represented by the gauge invariant operator matrix element
\cite{Becher:2010tm}
\begin{align}
  \label{eq:Bcal_quark}
  \Bcal_{q/N}(z,x_T^2) &= \frac{1}{2\pi} \int d t \, e^{-i z t \nb \cdot p} \, \sum_X \frac{\not\!\nb_{\alpha\beta}}{2} \, \braket{N(p) | \bar{\chi}_{\alpha}^n(t\nb+x_\perp) | X} \braket{X | \chi_\beta^n(0) | N(p)} \, , 
\end{align}
where the sum is over all intermediate states $X$ and the summation over the spinor indices $\alpha$, $\beta$ of the gauge invariant collinear quark field $\chi^n$ in SCET is understood.
The corresponding anti-quark TPDF is given by a similar equation with the role of the fields $\chi^n$ and $\bar{\chi}^n$ interchanged.
The TPDFs along the opposite direction, to which we refer as anti-collinear, are given by the same expressions, but with $p\sim n$ and $\bar{p} \sim \nb$ interchanged.
The regularization of the rapidity divergences which we will outline in section~\ref{sec:singularities} actually leads to a breaking of this relation. 
To mark this difference the anti-collinear TPDF will be denoted as $\bar{\Bcal}$.
In most aspects the discussion of these two different TPDFs is, however, completely analogous and for simplicity we therefore usually formulate it below only in terms of the collinear function.

For processes initiated by gluon-gluon fusion, factorization theorems similar to \eqn{eq:d3sigmadMdQtdY_sBsB} hold in which gluon TPDFs are encountered. 
Along the $n$ direction the latter is represented by the operator matrix element
\cite{Mulders:2000sh,Becher:2012yn}
\begin{align}
\label{eq:Bcal_gluon}
\Bcal_{g/N}^{\mu\nu}(z,x_\perp) &= \frac{-z \nb \!\cdot\! p}{2\pi} \!\int \!d t \, e^{-i z t \nb \cdot p} \sum_X \braket{N(p) | \Acal_{n,\perp}^{\mu a}(t\nb+x_\perp) | X}\! \braket{X | \Acal_{n,\perp}^{\nu a}(0) | N(p)} \, 
,
\end{align}
where $\Acal_{n,\perp}^{\mu a}$ is the gauge invariant collinear gluon field in SCET and the sum over the color index $a$ is understood.
Note that the gluon TPDF is a Lorentz tensor 
\cite{Mulders:2000sh,Catani:2010pd,Becher:2012yn}
in the space perpendicular to $n$ and $\nb$. It can be decomposed into two independent components as
\begin{align}
\label{eq:tensor_compo}
  \Bcal^{\mu\nu}_{g/N}(z,x_\perp) = \frac{g^{\mu\nu}_\perp}{d-2} \, \Bcal_{g/N}(z,x_T^2) + \left[ \frac{g^{\mu\nu}_\perp}{d-2} + \frac{x_\perp^\mu x_\perp^\nu}{x_T^2} \right] \Bcal'_{g/N}(z,x_T^2) \, ,
\end{align}
where $d$ is the number of space-time dimensions, $g^{\mu\nu}_\perp$ is the metric tensor in the transverse space and the projection onto the two components are given by
\begin{align}
  \Bcal_{g/N}(z,x_T^2) &= g_{\perp\mu\nu} \,  \Bcal^{\mu\nu}_{g/N}(z,x_\perp) \, , \nonumber
  \\
  \Bcal'_{g/N}(z,x_T^2) &= \frac{1}{d-3} \left[ g_{\perp\mu\nu} + (d-2) \, \frac{x_{\perp\mu} x_{\perp\nu}}{x_T^2} \right] \Bcal^{\mu\nu}_{g/N}(z,x_\perp) \, .
  \label{eq:sec_tensor_struct}
\end{align}

If the transverse scale is in the perturbative region, $x_T \ll 1/\Lambda_{QCD}$, the physics of these two scales can be factorized and the TPDFs can be matched onto the collinear PDFs defined as
\begin{align}
  \phi_{q/N}(z) &= \frac{1}{2\pi} \int d t \, e^{-i z t \nb \cdot p} \, \sum_X \frac{\not\!\nb_{\alpha\beta}}{2} \, \braket{N(p) | \bar{\chi}_{\alpha}^n(t\nb) | X} \braket{X | \chi_\beta^n(0) | N(p)} \, , \nonumber
  \\
  \label{eq:phi}
  \phi_{g/N}(z) &= -g_{\perp\mu\nu} \frac{z \nb \cdot p}{2\pi} \int d t \, e^{-i z t \nb \cdot p} \, \sum_X \braket{N(p) | \Acal_{n,\perp}^{\mu a}(t\nb) | X} \braket{X | \Acal_{n,\perp}^{\nu a}(0) | N(p)} \, .
\end{align}
These can be obtained from the TPDFs (\ref{eq:Bcal_quark}) and (\ref{eq:Bcal_gluon}) by setting $\xp=0$, corresponding to integrating over the transverse momentum. The matching takes the form \cite{Stewart:2010qs,Becher:2010tm,Becher:2012yn}
\begin{align}
\nonumber
  \Bcal_{i/N}(z,x_T^2) &= \sum_{j} \Ical_{i/j}(z,x_T^2) \otimes \phi_{j/N}(z) \, ,
  \\
  \Bcal_{g/N}'(z,x_T^2) &= \sum_{j} \Ical_{g/j}'(z,x_T^2) \otimes \phi_{j/N}(z) \, ,
\label{eq:match}
\end{align}
where the sum is over all partons $j$.
This holds up to power corrections in $x_T^2 \Lambda_{QCD}^2$ and we introduced the symbol $\otimes$ to denote the convolution of two functions as
\begin{align}
  f(z,\cdots) \otimes g(z,\cdots) \equiv \int_z^1 \frac{d\xi}{\xi} \, f(\xi,\cdots) \, g(z/\xi,\cdots) \, .
\end{align}
The matching kernels $\Ical_{i/j}$ and $\Ical_{g/j}'$ are perturbative functions. 
They can be extracted from \eqn{eq:match} and the perturbative parton-to-parton (T)PDFs $\Bcal_{i/j}$, $\Bcal_{g/j}'$ and $\phi_{i/j}$ given by eqs.~(\ref{eq:Bcal_quark},\,\ref{eq:Bcal_gluon},\,\ref{eq:phi}) with a parton $j$ in place of the hadron $N$.

With the results of the matching kernels and \eqn{eq:match}, the semi-perturbative hadron-to-parton TPDFs can be obtained from the collinear PDFs as long as $x_T$ is a perturbative scale.
The collinear PDFs have been extracted with high precision from experimental data by several groups. The knowledge of the matching kernels therefore provides an accurate determination of the TPDFs in the semi-perturbative domain. This not only has many phenomenological applications on its own right, but also provides necessary information for the determination of the fully non-perturbative TPDFs.

While $\Ical_{i/j}$ starts at $\as^0\,$, $\Ical_{g/j}'$ only starts at $\as^1$.
In many gluon-gluon initiated processes of interest, for example the production of a Higgs boson, the hard tensor contracting the Lorentz indices of two gluon TPDFs does not mix their two tensor structures.
Then for the same level of accuracy for physical observables,  the perturbative expansion of $\Ical_{g/j}'$ is required to one order less in $\as$ than the expansion of $\Ical_{g/j}$.  
In these cases the NLO expression of $\Ical_{g/j}'$ which was derived previously in \cite{Becher:2012yn} suffices for N$^3$LL precision.
For this reason, the main goal of this article is to determine the NNLO corrections to the matching kernels $\Ical_{i/j}$ from those of the parton-to-parton (T)PDFs, while  $\Ical'_{g/j}$ at this order will be discussed in a forthcoming article.

\subsection{Treatment of singularities}
\label{sec:singularities}

In the calculation of $\Bcal_{i/j}$ we have to deal with several kinds of singularities.
On the one hand there are the usual ultraviolet (UV) and infrared (IR) singularities, which we regulate by dimensional regularization in $d=4-2\epsilon$ dimensions.
On the other hand the functions contain extra light-cone singularities which require additional regularization.
As mentioned in the introduction, there are several proposals of regulators. 
In our calculation, we use the analytic regulator as suggested in \cite{Becher:2011dz}. This amounts to introducing in the phase space integrals a factor $( \nu / n\cdot l_i )^\alpha$ for each unresolved final-state parton with momentum $l_i$. Here $\alpha$ is the analytic regulator and $\nu$ is an unphysical mass scale associated with the regulator --- in a similar way as the renormalization scale $\mu$ is related to the dimensional regulator $\epsilon$.
Note that the regulating factor has to contain the same light-cone component $n\cdot l_i$ for both the collinear and the anti-collinear region.
As such, it breaks the symmetry $p\sim n \leftrightarrow \bar{p} \sim \nb$ between the two regions and the rescaling invariance of SCET, a fact called ``collinear anomaly'' in \cite{Becher:2010tm}. These symmetries will be restored at the end of the calculation when all divergences are removed and the limit $\alpha \to 0$ is taken.
One good property of this scheme is that the soft function $\cal{S}$ (for Drell-Yan like processes) automatically reduces to a trivial factor of unity. For other regulators where this is not the case, one may always absorb the soft function into the two TPDFs by a redefinition (see, e.g., \cite{Collins:2011zzd}).

The analytic regulator combined with the dimensional one suffices to regulate all singularities in the operator matrix elements.
The singularities manifest themselves in the TPDFs as poles in the regulators.
While the individual factors $\cal{S}$, $\Bcal$ and $\bar{\Bcal}$ are scheme dependent, their product is well defined and especially all poles in the regulator $\alpha$ cancel therein, along with the dependence on the unphysical scale $\nu$ after the limit $\alpha \rightarrow 0$ is taken.
However, a dependence on the hard scale $q^2$ remains even after the regulator is dropped.
The generation of the hard scale $q^2 \sim (\nb \cdot p)(n \cdot \bar{p})$ through the analytic regulator can be understood from the scale ratio $(\nu / n \cdot l_i)$ appearing along with it.
For the anti-collinear region this ratio can be expressed in terms of $(\nu/n \cdot \bar{p})$, while for the collinear region it can be expressed in terms of $(\nu \, \nb \cdot p \, x_T^2)$. 
In the combination of the two factors, the scale $\nu$ drops out, while the mass ratio $q^2 x_T^2$ is left over.

Extending these arguments, using the existence of the $\alpha\rightarrow 0$ limit of the product of two corresponding TPDFs and its independence on the scale $\nu$, it was shown in \cite{Becher:2010tm} that the product can be refactorized into the form
\begin{align}
  \label{eq:refactorization} 
  \lim_{\alpha \to 0} \left[ \mathcal{S}(x_T^2) \Bcal_{i/ j}(z_1,x_T^2) \bar{\Bcal}_{\ib /k}(z_2,x_T^2) \right]_{q^2} &= \nonumber
  \\
  &\hspace{-3.5cm}
  \left( \frac{x_T^2 q^2}{4e^{-2\gamma_E}} \right)^{-F^b_{i\ib }(x_T^2)}
  B^b_{i/j}(z_1,x_T^2) \, B^b_{\ib /k}(z_2,x_T^2)
  \,,
\end{align}
where after the cancellation of all poles in $\alpha$ on the left hand side the analytic regulator is set to zero and the right hand side is free of both $\alpha$ and $\nu$.
This defines the anomaly coefficient $F$ and the true TPDFs $B_{i/j}$ which are universal process-independent functions and  have the same form for the collinear and anti-collinear region.

These functions still contain poles in $\epsilon$ as indicated by the label $b$ for bare. 
By the operator renormalization
\begin{align}
\label{eq:renorm_B}
B_{i/j}^{b}(z,x_T^2) &= Z^B_i(x_T^2,\mu) \, B_{i/j}(z,x_T^2,\mu) \, , 
\\
\label{eq:renorm_F}
F^{b}_{i\ib }(x_T^2) &= F_{i\ib }(x_T^2,\mu) + Z^F_i(\mu) \, ,
\end{align}
the UV poles are absorbed into the renormalization factors $Z$, such that the renormalized functions $B_{i/j}(z,x_T^2,\mu)$ and $F_{i\ib }(x_T^2,\mu)$ are free of these singularities.
Upon renormalization a dependence on the renormalization scale $\mu$ is introduced which is described by the renormalization group equations (RGEs) and will be discussed further below.
We work in the $\overline{\text{MS}}$ scheme, which amounts to expressing the bare coupling constant as
\begin{align}
\label{eq:as_bare}
\as^{b} = \left(\frac{\mu^2 e^{\gamma_E}}{4\pi}\right)^{\epsilon} Z_{\alpha}(\mu)  \as(\mu)
\,,
\end{align}
and requiring that the renormalization factors contain only poles in $\epsilon$.
After renormalization, $F$ is free of any poles, while $B_{i/j}$ can still contain IR poles. This signals the non-perturbative nature of the TPDFs. 
These IR poles are exactly the same as those in the collinear PDFs, whose renormalization takes the form
\begin{align}
  \label{eq:renorm_phi}
  \phi_{i/j}^{b}(z) &= \sum_{k} Z^\phi_{i/k}(z,\mu) \otimes \phi_{k/j}(z,\mu) \, .
\end{align}
Just as the functions $\Bcal_{i/j}$ and $\phi^b_{i/j}$ are related by \eqn{eq:match}, the transverse and collinear PDFs are related by matching kernels via
\begin{align}
\label{eq:match_reg}
  B_{i/j}(z,x_T^2,\mu) &= \sum_{k} I_{i/k}(z,x_T^2,\mu) \otimes \phi_{k/j}(z,\mu) \, ,
\end{align}
in both their renormalized and bare versions.
This relation, the renormalization of $B$ and $\phi$ as well as the result
\begin{align}
\label{eq:phi_bare}
\phi_{i/j}^{b}(z) = \delta_{ij}\delta(1-z)\,,
\end{align}
to all orders in dimensional regularization
imply the renormalization of the matching kernels as
\begin{align}
  I^{b}_{i/j}(z,x_T^2) &= Z^{B}_i(x_T^2,\mu)\sum_k I_{i/k}(z,x_T^2,\mu)\otimes \phi_{k/j}(z,\mu) \, .
  \label{eq:renorm_I}
\end{align}
In this equation the UV poles are contained in $Z^{B}_i$ and the IR poles in $\phi_{k/j}$, while $I_{i/k}$ is free of any poles.
In fact even though we do not explicitly distinguish IR and UV poles in our calculation, this equation allows us not only to extract the renormalized matching kernel, but also separately the renormalization factor $Z^{B}_i$ and the renormalized PDFs.
The separation of the last two functions can be achieved by fixing the endpoint contributions of the renormalized PDFs $\phi_{j/j}$ from constraints on their integrals implied from momentum and quark number conservation.

\subsection{Resummation}
The differential cross section \eqref{eq:d3sigmadMdQtdY_sBsB} can now be written as \cite{Becher:2010tm}
\begin{align}
\label{eq:dsigma_DY_full}
\frac{d^3\sigma}{dq^2dq_T^2 d y} 
=\;& \frac{\alpha^2}{3N_c q^2 s} 
 \sum_{i,j} 
\sum_{q}
 e_q^2
\left[
C_{q\bar{q}\leftarrow ij}
( z_1,z_2,q_T^2,q^2,\mu) + ( q\leftrightarrow \bar{q} ) \right]
\nonumber\\
&\hspace{3.1cm}\otimes\phi_{i/N_1}( z_1,\mu)\otimes\phi_{j/N_2}( z_2,\mu)
\,,
\end{align}
which holds up to power corrections in $q_T^2/M^2$ and $x_T^2 \Lambda_\text{QCD}^2$ with the perturbative function
\begin{align}
C_{q\bar{q}\leftarrow ij}
( z_1,z_2,q_T^2,q^2,\mu)
=\,&
\left| C_V( -q^2,\mu)\right|^2 \!
\int\!\!  d^2 \! x_\perp\, e^{-i q_\perp\!\cdot x_\perp} 
\left(\frac{x_T^2q^2}{4e^{-2\gamma_E}} \right)^{\!\!\!-F_{q\bar{q}}(x_T^2,\mu)}
\nonumber\\
& \times
I_{q/ i}( z_1,x_T^2,\mu) I_{\bar{q}/ j}( z_2,x_T^2,\mu) \,.
\label{eq:Cqqb_DY}
\end{align}
The functions appearing here can be related to the quantities $A$, $B$ and $C_{ij}$ as defined in \cite{Collins:1984kg}. 
These relations are given in eqs.~(71,\,72) of \cite{Becher:2010tm}.

In \eqn{eq:Cqqb_DY} each function depends only on a single physical mass scale and can be determined consistently in fixed order perturbation theory by choosing the scale $\mu$ in the vicinity of that scale such that no large logarithms are present.
In a subsequent step, all functions have to be matched at the same scale $\mu$. 
This is achieved by solving the RGEs for each of them, which automatically resums all large logarithms.

In terms of the logarithm
\begin{align}
\label{eq:Lp}
L_\perp = \log{\frac{x_T^2\mu^2}{4e^{-2\gamma_E}}}
\,,
\end{align}
the DGLAP splitting kernels $P_{j k}(z)$, the cusp anomalous dimension in the fundamental (adjoint) representation $\Gamma^q_\text{cusp}$ ($\Gamma^g_\text{cusp}$) and the quark (gluon) anomalous dimension $\gamma^q$ ($\gamma^g$), which are all listed in appendix \ref{sec:ad}, the RGEs can be written as
\begin{align}
\label{eq:RGE_CV} 
\frac{d}{d\log \mu}C_V(-q^2,\mu)
=
&\left[\Gamma^q_{\mathrm{cusp}}(\as) \log{\frac{-q^2}{\mu^2}}
+2\gamma^q(\as)\right]C_V(-q^2,\mu)
\,,
\displaybreak[1]
\\
\label{eq:RGE_F}
\frac{d}{d\log\mu}F_{i\ib }(x_T^2,\mu)=&\;2\,\Gamma^i_{\mathrm{cusp}}(\as)
\,,
\displaybreak[1]
\\
\frac{d}{d\log \mu}I_{i/j}(z,x^2_T,\mu) = &
\Big[
\Gamma^i_{\text{cusp}}(\alpha_s) L_\perp - 2\gamma^i(\alpha_s)
\Big] I_{i/j}(z,x^2_T,\mu) 
\nonumber\\&
- 2 \sum_k I_{i/k}(z,x_T^2,\mu)\otimes P_{k j}(z,\mu)
\label{eq:RGE_I}
\,.
\end{align}
The last equation follows from the RGEs of the (T)PDFs
\begin{align}
\label{eq:RGE_B}
\frac{d}{d\log \mu}B_{i/N}(z,x_\perp^2,\mu)
=
&\left[\Gamma^i_{\mathrm{cusp}}(\as) L_\perp 
-2\gamma^i(\as)\right]B_{i/N}(z,x_\perp^2,\mu)
\,,
\\
\label{eq:DGLAP}
\frac{d}{d\log \mu}\phi_{j/N}(z,\mu)
=&
\,2 \sum_k P_{j k}(z,\mu) \otimes \phi_{k/N}(z,\mu)
\,,
\end{align}
and \eqn{eq:match_reg}.
Eq.~\eqref{eq:RGE_CV} takes a corresponding form for other processes; for gluon initiated processes with the anomalous dimensions $\Gamma^g_{\mathrm{cusp}}$ and $\gamma^g$.
This equation and the independence of the cross section on $\mu$ imply eqs.~(\ref{eq:RGE_F},\,\ref{eq:RGE_B}). 
Also note the appearance of the hard scale $q^2$ in \eqn{eq:RGE_CV} already implied a compensating dependence on this scale for the other factors. This has been found in terms of the collinear anomaly, \eqn{eq:refactorization}.

Since the bare functions do not depend on $\mu$, each renormalization constant in eqs.~(\ref{eq:renorm_B} -- \ref{eq:renorm_phi}) obeys a RGE which exactly compensates the $\mu$ dependence of the corresponding renormalized function. 
Solving these equations and enforcing the \msbar\ condition, which is most conveniently done using the $d$ dimensional coupling constant, allows us to express the renormalization constants in terms of the corresponding anomalous dimensions and the QCD $\beta$ function.
The results for $\phi_{i/j}$, $Z^B_i$ and $Z^F_i$ are listed in appendix~\ref{sec:ren_factors}.
Comparing these expectations with the findings in our calculation serves as  a check on our results.

Provided that all coefficients in eqs.~(\ref{eq:Cqqb_DY} -- \ref{eq:DGLAP}) and the QCD $\beta$ function are determined to sufficient order, any logarithmic precision goal for the differential cross section can be achieved.
To obtain e.g.\ N$^3$LL precision, the Wilson coefficient and $I_{i/j}$ have to be known to $\as^2$, while $F_{i\ib}$, $P_{k j}$ and $\gamma^i $ are needed to $\as^3$. 
Moreover, $\Gamma_\text{cusp}$ and $\beta$ are needed to $\as^4$. 
Only some of them are known to this accuracy, which are $\beta$ in \cite{vanRitbergen:1997va}, $P_{k j}$ in \cite{Moch:2004pa,Vogt:2004mw}, $\gamma^i$ in \cite{Becher:2009qa} and for several processes also the Wilson coefficients.

The derivation of the $\as^2$ contributions to  $I_{i/j}$, as required for the N$^3$LL transverse momentum resummation, is the main objective of this paper.
These and $F_{i\ib}$ up to $ \as^2$ can be obtained in the way outlined in this section from a perturbative calculation of $\Bcal_{i/j}$, $\bar{\Bcal}_{i/j}$ and $\phi^{b}_{i/j}$ up to NNLO in $\as$, i.e.\ the expansion of 
\begin{align}
\label{eq:pert_expansion}
f(\as,\ldots) = \sum_{n=0}^\infty \left(\asFPi\right)^n f^{(n)}(\ldots)
\end{align}
up to $n=2$. This calculation is discussed below.
The main results, the NNLO matching kernels $I^{(2)}_{i/j}$, are presented in section~\ref{sec:results}. 
For completeness we also list further relevant perturbative results in appendix~\ref{sec:outsourced_results}.

\section{Perturbative calculation}
\label{sec:pert_calc}
Once the perturbative results for the parton-to-parton (T)PDFs to sufficient order in the strong coupling and the two regulators are determined, the extraction of the final results according to eqs.~(\ref{eq:refactorization} -- \ref{eq:renorm_I}) is straightforward.
We therefore only discuss the former in more detail. We begin with the collinear case.

Since the relevant matrix elements (\ref{eq:Bcal_quark},\,\ref{eq:Bcal_gluon},\,\ref{eq:phi}) contain solely collinear fields and the purely collinear SCET Lagrangian has the same form as the full QCD Lagrangian, we can use QCD Feynman rules to evaluate them.
In a general gauge, any number of gluons can couple to the Wilson lines contained in the gauge invariant fields ($\chi$, $\cal{A}$) and the associated vertices lead to denominators with momentum components projected to the $\nb$ direction.
A special gauge is the light cone gauge with $\nb$ chosen as the light cone vector. 
In this gauge the Wilson lines reduce to factors of unity, but one still finds the $\nb$ dependent denominators --- this time introduced through the gluon propagators.
We will focus our discussion to this gauge, although we also performed the calculation in Feynman gauge as a cross check.

In our regularization scheme, the perturbative corrections to the bare collinear PDFs lead to scaleless integrals vanishing in dimensional regularization, such that their all order results are given by \eqn{eq:phi_bare}.

\begin{figure}[tp]
\centering
  \begin{subfigure}[b]{0.22\textwidth}
   \includegraphics[width=\textwidth]{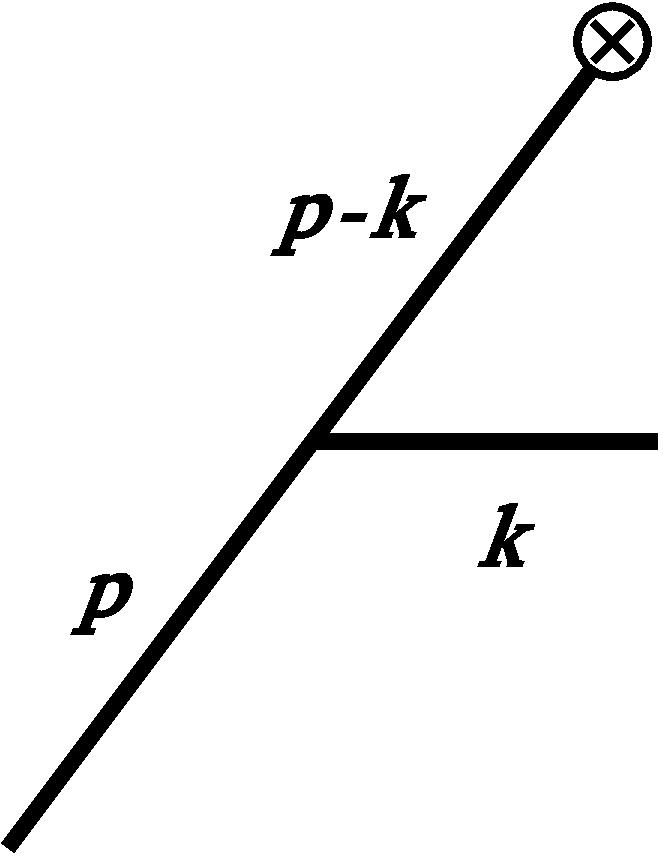}
   \caption{ }
  \end{subfigure}
\hspace{0.8cm}
  \begin{subfigure}[b]{0.25\textwidth}
   \includegraphics[width=\textwidth]{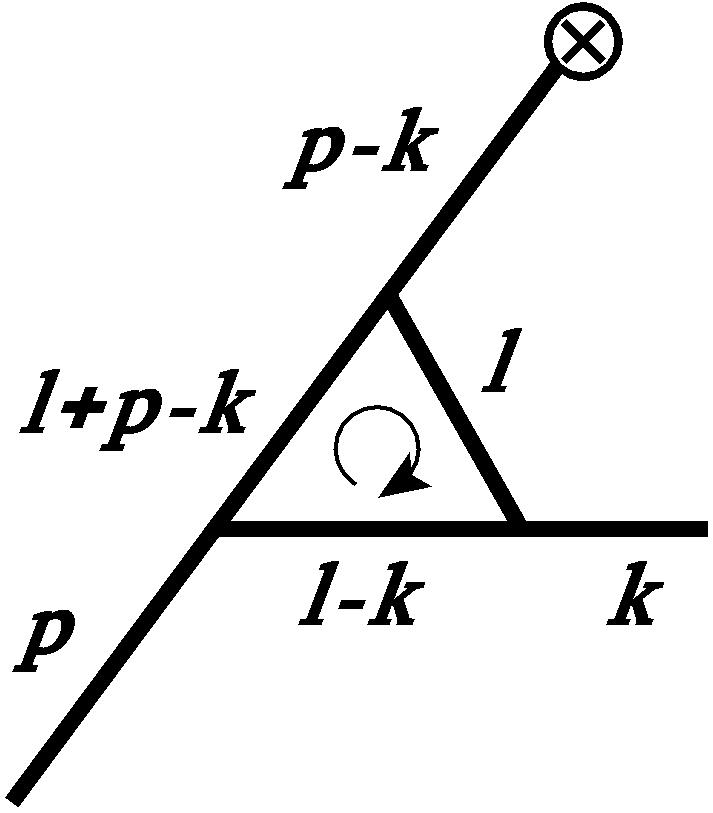}
   \caption{ }
  \end{subfigure}
\hspace{0.8cm}
  \begin{subfigure}[b]{0.21\textwidth}
   \includegraphics[width=\textwidth]{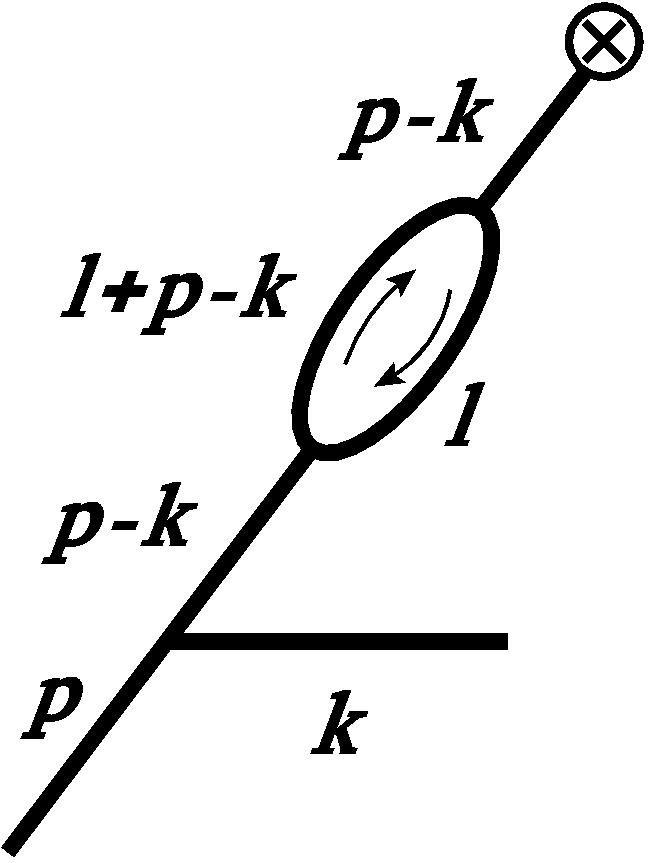}
   \caption{ }
  \end{subfigure}
 \caption{ Amplitude topologies for real (a) and virtual-real (b,c) case.}
 \label{fig:topology_VR}
 \end{figure}

\begin{figure}[tp]
\centering
  \begin{subfigure}[b]{0.23\textwidth}
   \includegraphics[width=\textwidth]{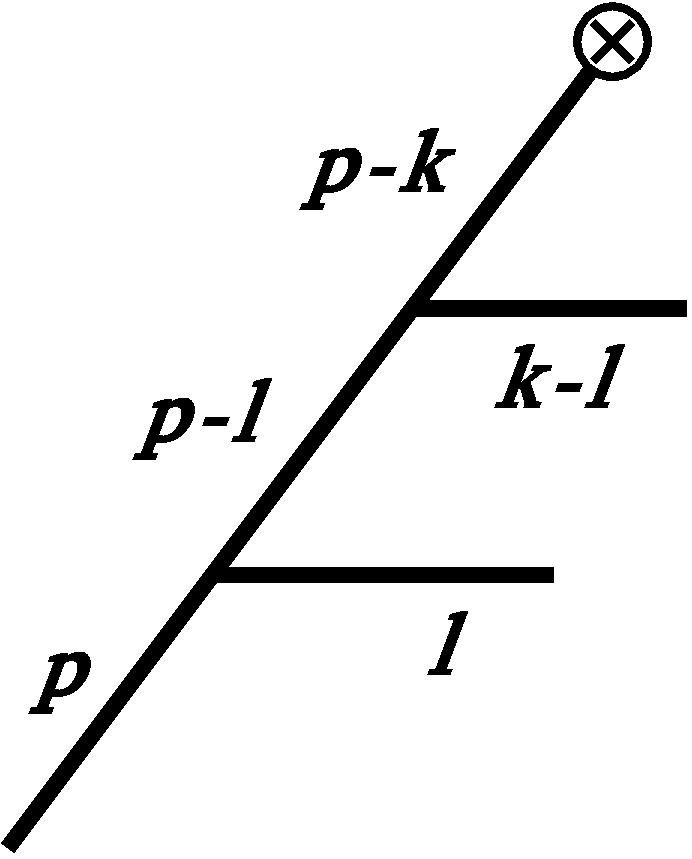}
   \caption{ }
   \end{subfigure}
\hspace{0.8cm}
  \begin{subfigure}[b]{0.23\textwidth}
   \includegraphics[width=\textwidth]{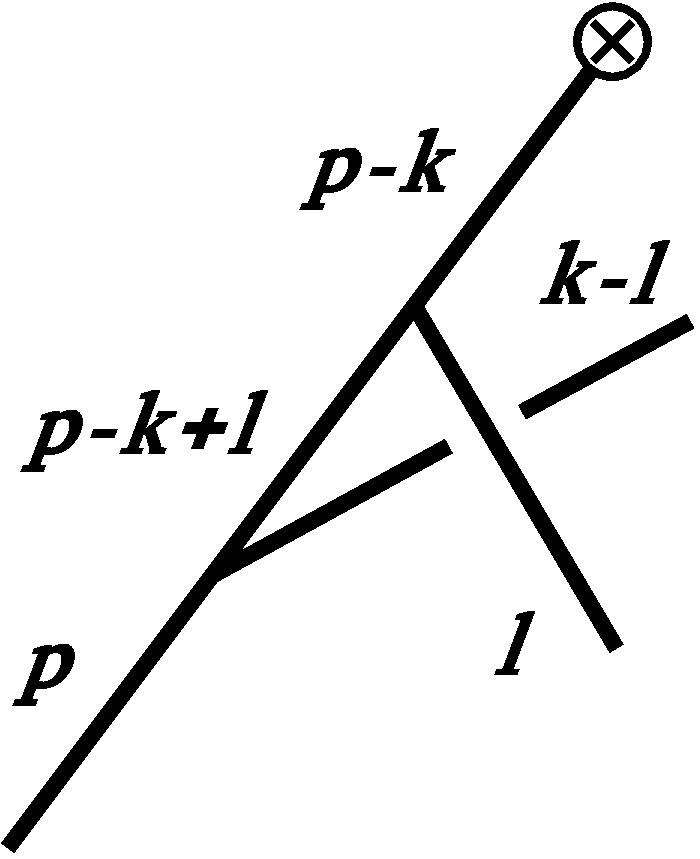}
   \caption{ }
  \end{subfigure}
\hspace{0.8cm}
  \begin{subfigure}[b]{0.25\textwidth}
   \includegraphics[width=\textwidth]{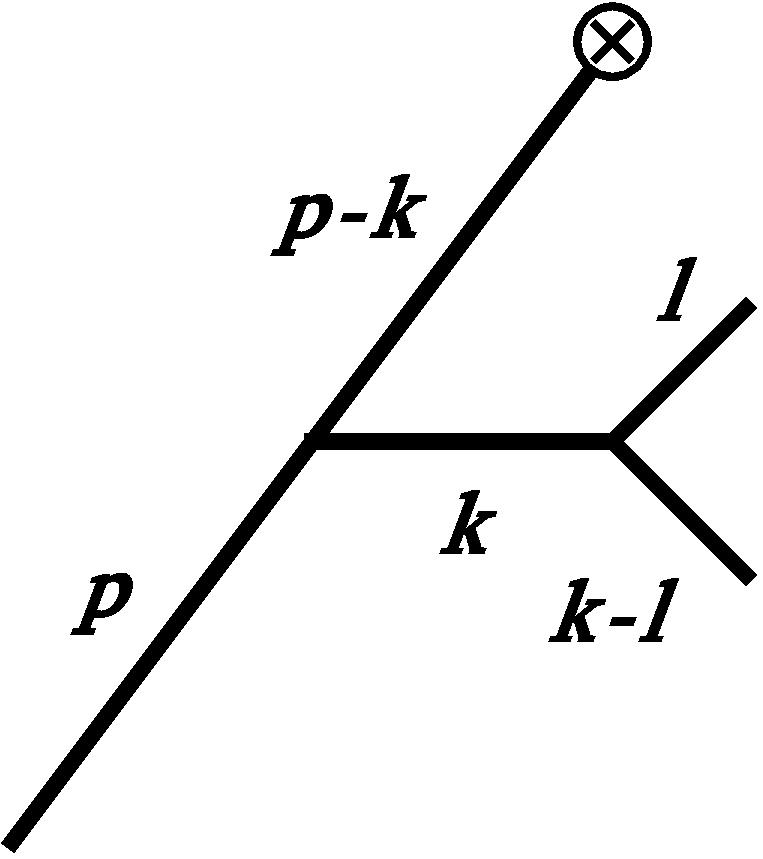}
   \caption{ }
  \end{subfigure}
 \caption{ Amplitude topologies for double real case.}
\label{fig:topology_RR}
\end{figure}

For the transverse PDFs the additional scale $x_\perp$ is present.
The corresponding expressions essentially correspond to the square of matrix elements as in figures~\ref{fig:topology_VR} and \ref{fig:topology_RR} where the momentum $p-k$ of the parton coupling to the gauge invariant field can be off-shell.
Calling the momenta of the emitted partons $l_i$, $i=1,\ldots n_r$, and their sum $k=\sum_i l_i$, the phase space factor takes the form
\begin{align}
\label{eq:PS_TD}
\int\!\!d \Pi_{n_r}^\text{TD}
&=
\left[\prod_{i}\int\! \frac{d^d l_i}{(2\pi)^{d-1}}\,\delta^+\!\big(l_i^2\big)  \left( \frac{\nu}{n\cdot l_i} \right)^\alpha \right]
\int \! d^d k
\;\delta^d\Big(k-\sum_{i}l_i\Big)
e^{-ik_\perp \cdot x_\perp} 
\delta\big(\hat{k}_z\big)
\,,
\end{align}
where $ \hat{k}_z = \nb\tcdot  [k - (1-z) p]$.
The last factor arose from the $t$ integral in eqs.~(\ref{eq:Bcal_quark},\,\ref{eq:Bcal_gluon}), the exponential from the $\xp$ dependence of the gauge invariant fields and the $\alpha$ dependent factors arise from the analytic regularization.
It is essentially these factors which lead to difficult integrals, where many standard  calculational methods become inapplicable.
Another complication is the presence of light-cone propagators due to the use of light-cone gauge (or alternatively the presence of Wilson lines).

For the anti-collinear case, the arguments are completely analogous. Relabeling $p\sim n \leftrightarrow \bar{p}\sim \nb$, one finds the same expressions as for the collinear cases, the only change is the appearance of the analytic regulator which now enters in \eqn{eq:PS_TD} as $(\nu / \nb \cdot l_i)^\alpha$.
Using this relabeling, in the following we can discuss the collinear and anti-collinear cases in parallel.

Discussing the individual contributions up to NNLO, we first observe that for $n_r=0$ emitted partons, $k=0$ and the $\xp$ dependence is lost.
Setting $\alpha = 0$, no scale dependence remains in these cases and dimensionless integrals are found.
Hence, the bare TPDFs receive no contributions from purely virtual corrections.
Then to obtain the corrections up to $\as^2$ to the trivial LO results $\Bcal^{(0)}_{i/j}(z,x_T^2),\bar{\Bcal}^{(0)}_{i/j}(z,x_T^2) = \delta_{ij}\delta(1-z)$, the only cases we have to consider are the real NLO corrections as well as the double real and the virtual real NNLO corrections.
Their amplitudes correspond to the diagrams in figures~\ref{fig:topology_VR} and \ref{fig:topology_RR} with appropriate placement of partons as well as diagrams obtained from shrinking individual lines to points.
The full contributions are obtained from appropriate combinations of these amplitudes with their Hermitian conjugates. Sums and averages over color and spin of external partons are understood.
Therein the factors in eqs.~(\ref{eq:Bcal_quark},\,\ref{eq:Bcal_gluon},\,\ref{eq:sec_tensor_struct}) contracting the two gauge invariant fields lead to the factor $\nb_{\alpha\beta} / 2$ if they are (anti)-quarks, and $-z\nb\tcdot p \, g_{\perp\mu\nu}$ if they are gluons.

In this sense, the NLO contributions correspond to the square of diagram \ref{fig:topology_VR}(a).
The two different 1-loop amplitude topologies with unspecified partons are depicted in figure~\ref{fig:topology_VR}(b,c). 
For the virtual-real contribution, these diagrams or their versions with a shrinked propagator are combined with the NLO diagram of figure~\ref{fig:topology_VR}(a).
The double real diagrams without specified partons are given in figure~\ref{fig:topology_RR}. 
For all three amplitude topologies one propagator carries momentum $p-k$. The other momentum is either $p-l$, $p-k+l$ or $k$ depending on the amplitude topology.
By shrinking the propagator with the second momentum, we receive the same additional amplitude subtopology from all of them. 
For the double real NNLO contribution, these diagrams are combined with each other.
We use QGRAF \cite{Nogueira:1991ex} to generate the amplitudes and FORM \cite{Vermaseren:2000nd} to manipulate them.

The NLO contributions can be solved in closed form.
Having used the $\delta$ distributions, the only integral required is
\begin{align}
  \label{eq:intkt}
  &\frac{\mu^{2\epsilon+2\delta}}{\pi^{1-\epsilon}} \int \frac{d^{2-2\epsilon}k_T}{k_T^{2+2\delta}} \, e^{ik_T \cdot x_T} = e^{-2(\epsilon+\delta)\gamma_E} \, \frac{\Gamma(-\epsilon-\delta)}{\Gamma(1+\delta)} \left(\frac{x_T^2\mu^2}{4e^{-2\gamma_E}} \right)^{\epsilon+\delta} \!\!\raisebox{-5pt}{.} 
\end{align}
The corresponding results are given in \eqn{eq:sB_AC_nlo}.
Using appropriate pa\-ra\-me\-tri\-za\-tions, we will identify this integral as subset of the integrals of the two NNLO cases --- the virtual-real and double real corrections, which we discuss in the following two subsections.

Expressing the bare coupling constant via \eqn{eq:as_bare} by the renormalized one,
introduces powers of the \msbar\ factor and 
an additional NNLO contribution stemming from the NLO contribution multiplied by the $\as^1$ term in the renormalization factor
\begin{align}
  Z_\alpha = 1 + \frac{\alpha_s}{4\pi} \left( -\frac{\beta_0}{\epsilon} \right) + \cdots \, .
\end{align}
\subsection{Virtual-real contribution}
The calculation of the virtual-real diagrams is straightforward. We first perform the integrals over the loop momenta. Using partial fraction decomposition and shift of momentum, we can reduce the scalar loops integrals to two generic types:
\begin{align}
  I^{\text{VR}}_1(a_1,a_2,a_3,a_4) &= \int \frac{d^d l}{(2\pi)^d} \left[ -l^2 \right]^{-a_1} \left[ -(l+q)^2 \right]^{-a_2} \left[ -(l+p)^2 \right]^{-a_3} \left[ \nb \cdot l \right]^{-a_4} , \nonumber
  \\
  I^{\text{VR}}_2(a_1,a_2,a_3,a_4) &= \int \frac{d^d l}{(2\pi)^d} \left[ -l^2 \right]^{-a_1} \left[ -(l+q)^2 \right]^{-a_2} \left[ -(l-k)^2 \right]^{-a_3} \left[ \nb \cdot l \right]^{-a_4} ,
\end{align}
where $q=p-k$ and in all the propagators an imaginary part of $-i\delta$ is implicit. These integrals can be calculated using standard techniques. Taking $I^{\text{VR}}_1$ as an example, we first use a Feynman parameterization to combine the propagators and perform the integration over the loop momentum. We are then left with a multi-dimensional integral over the Feynman parameters:
\begin{align}
  I^{\text{VR}}_1(a_1,a_2,a_3,a_4) &= \frac{i}{2^{4-2\epsilon}\pi^{2-\epsilon}} \cdot\frac{\Gamma(a_1+a_2+a_3+a_4-2+\epsilon)}{\Gamma(a_1) \, \Gamma(a_2) \, \Gamma(a_3) \, \Gamma(a_4)} \nonumber
  \\
  &\hspace{-6em} \times \int_0^1 [d x] \int_0^\infty d\lambda \, x_1^{a_1-1} \, x_2^{a_2-1} \, x_3^{a_3-1} \, \lambda^{a_4-1}
(x_1+x_2+x_3)^{a_1+a_2+a_3+a_4-4+2\epsilon}
 \nonumber
  \\
  &\hspace{-6em} \times  \left[ -q^2 x_1 x_2 - \nb \cdot (x_2 q + x_3 p) \, \lambda \right]^{2-\epsilon-a_1-a_2-a_3-a_4} \, , 
\end{align}
where $[d x] = d x_1 \, d x_2 \, d x_3 \, \delta(x_1+x_2+x_3-1)$. The remaining integrals are not difficult to carry out and the results can be written in closed form in terms of hypergeometric functions. The final forms of the two integrals are
\begin{align}
  I^{\text{VR}}_1(a_1,a_2,a_3,a_4) &= \frac{i}{2^{4-2\epsilon}\pi^{2-\epsilon}} \left( -q^2 \right)^{2-\epsilon-a_1-a_2-a_3} \left( -\nb\tcdot p -i\delta \right)^{-a_4} \nonumber
  \\
  &\hspace{-5em} \times \frac{\Gamma(a_1+a_2+a_3-2+\epsilon) \, \Gamma(2-\epsilon-a_1-a_3) \, \Gamma(2-\epsilon-a_2-a_3) \, \Gamma(2-\epsilon-a_1-a_4)}{\Gamma(a_1) \, \Gamma(a_2) \, \Gamma(2-\epsilon-a_1) \, \Gamma(4-2\epsilon-a_1-a_2-a_3-a_4)} \nonumber
  \\
  &\hspace{-5em} \times {}_2F_1 \left( a_4, 2-\epsilon-a_1-a_3; 2-\epsilon-a_1; 1-z \right) ,
  \displaybreak[1]
  \\
  I^{\text{VR}}_2(a_1,a_2,a_3,a_4) &= \frac{i}{2^{4-2\epsilon}\pi^{2-\epsilon}} \left( -q^2 \right)^{2-\epsilon-a_1-a_2-a_3} \left( \nb\tcdot k \right)^{-a_4} \nonumber
  \\
  &\hspace{-5em} \times \frac{\Gamma(a_1+a_2+a_3-2+\epsilon) \, \Gamma(2-\epsilon-a_1-a_3) \, \Gamma(2-\epsilon-a_2-a_3) \, \Gamma(2-\epsilon-a_1-a_4)}{\Gamma(a_1) \, \Gamma(a_2) \, \Gamma(2-\epsilon-a_1) \, \Gamma(4-2\epsilon-a_1-a_2-a_3-a_4)} \nonumber
  \\
  &\hspace{-5em} \times {}_2F_1 \left( a_4, 2-\epsilon-a_1-a_3; 2-\epsilon-a_1; \frac{1}{1-z} \right) .
\end{align}
Having performed the loop integrals, the remaining integrals over $k$ are similar to those at NLO and can be readily evaluated using \eqn{eq:intkt}.

\subsection{Double real contribution}
\label{sec:2R}
For the combined double real emission diagrams we find integrals of the form
\begin{align}
  \label{eq:IRRmomenta}
  &\int d^{2-2\epsilon}k_T \, e^{ik_T \cdot x_T} \int_{k_T^2/\nb \cdot k}^\infty \!\!\!\!d(n\tcdot k) \int \!d^d l \, \delta^+(l^2) \, \delta^+((k-l)^2)
  \\
  &\hspace{3em} \times \overline{|M|}^2\! \left( \nb \cdot l, n \cdot l, \nb \cdot (k-l), n \cdot (k-l), k^2, -(p-k)^2, \nb \cdot (p-l), \nb \cdot (p-k+l) \right) , \nonumber
\end{align}
where the $\nb\tcdot k$ integral has already been performed using $\delta(\hat{k}_z)$ such that $\nb\tcdot k =(1-z)\nb\tcdot p\,$. 
In the argument of the squared amplitude $\overline{|M|}^2$ we have listed all possible scalar products that can appear.

We introduce a variable change $y=k_T^2/(n\tcdot k \, \nb \tcdot k)$, and the $n\tcdot k$ integral becomes
\begin{align}
  \int_{k_T^2/\nb\cdot k}^\infty \!\!\!\!d(n\cdot k) = \frac{k_T^2}{\nb \tcdot k} \int_0^1 \frac{d y}{y^2} \, .
\end{align}
To evaluate the $l$ integral, we boost into the rest frame of $k$, such that the vectors can be parameterized as
\begin{align}
  k^\mu &= k_T \sqrt{\frac{1-y}{y}} \, (1,\cdots,0,0,0) \, , \nonumber
  \displaybreak[1]\\
  \nb^\mu &= \frac{\nb \tcdot k}{k_T} \sqrt{\frac{y}{1-y}} \, (1,\cdots,0,0,1) \, , \nonumber
  \displaybreak[1]\\
  n^\mu &= \frac{k_T}{\nb \tcdot k} \frac{1}{\sqrt{y(1-y)}} \left( 1,\cdots,-2\sqrt{y(1-y)},2y-1 \right) , \nonumber
  \displaybreak[1]\\
  l^\mu &= \frac{k_T}{2} \sqrt{\frac{1-y}{y}} \, (1,\cdots,\sin\theta_1\sin\theta_2,\sin\theta_1\cos\theta_2,\cos\theta_1) \, ,
\end{align}
and the scalar products are given by
\begin{align}
  &\nb \cdot l = \frac{\nb \tcdot k}{2} \, (1-\cos\theta_1) \equiv \nb \tcdot k \, D_1 \, , \nonumber
  \\
  &n \cdot l = \frac{k_T^2}{2y\, \nb \tcdot k} \left[ 1 + 2\sqrt{y(1-y)} \sin\theta_1\cos\theta_2 - (2y-1) \cos\theta_1 \right] \equiv \frac{k_T^2}{y\,\nb \tcdot k} \, D_2 
  \, , \displaybreak[1]\nonumber
  \\
  &\nb \cdot (k-l) = \nb \tcdot k \, (1-D_1) \equiv \nb \tcdot k \, D_3 \, , \nonumber
  \\
  &n \cdot (k-l) = \frac{k_T^2}{y\, \nb \tcdot k} \, (1-D_2) \equiv \frac{k_T^2}{y\, \nb \tcdot k} \, D_4 
  \, , \displaybreak[1]\nonumber
  \\
  &k^2 = k_T^2 \, \frac{1-y}{y} \, , \nonumber
  \\
  &-(p-k)^2 = \frac{k_T^2}{y(1-z)} \, [1-(1-y)(1-z)] \equiv \frac{k_T^2}{y(1-z)} \, D_7 
  \, , \displaybreak[1]\nonumber \\
  &\nb \cdot (p-l) = \nb \tcdot p \, [1-(1-z)D_1] \equiv \nb \tcdot p \, D_8 \, , \nonumber
  \\
  &\nb \cdot (p-k+l) = \nb \tcdot p \, [1-(1-z)D_3] \equiv \nb \tcdot p \, D_9 \, .
\end{align}
We also define $D_5=y$ and $D_6=1-y$. From the above equations, we see that whenever $D_8$ and $D_9$ both appear, we can use a partial fraction decomposition to get rid of one of them. We can also use partial fraction decompositions for the pairs $\{D_1,D_3\}$ and $\{D_2,D_4\}$. However, it is not always possible to get rid of these due to the analytic regulator.

Inserting the above parameterizations into \eqn{eq:IRRmomenta}, we see that the $k_T$ dependence is power-like, and the $k_T$ integral can be easily performed using \eqn{eq:intkt}. Performing also the $l^0$ and $|\vec{l}|$ integrals using the delta functions, we finally arrive at integrals of the form
\begin{align}
  \label{eq:IRR}
  I^{\text{RR}}(\{a_i\}) &= \frac{1}{2\pi} \, \frac{\Gamma^2(1-\epsilon)}{\Gamma(1-2\epsilon)} \int_0^1 \!d y \int_0^\pi \!\!d\theta_1\, \sin^{1-2\epsilon}\!\theta_1 \!\int_0^\pi  \!\!d\theta_2\, \sin^{-2\epsilon}\!\theta_2 \, D_5^{\epsilon} \, D_6^{-\epsilon}  \prod_i D_i^{-a_i} \, ,
\end{align}
where $\{a_i\}=\{a_1,a_2,a_3,a_4,a_5,a_6,a_7,a_8,a_9\}$ is the collection of the powers of denominators. Either $\{a_1,a_3\}$ or $\{a_2,a_4,a_5\}$ will contain the analytic regulator $\alpha$, and in general we will calculate the integrals as a power series in $\alpha$ and $\epsilon$.

There are two situations where we need to keep the $\alpha$ regulator in the integrals. One is if the integral itself is divergent for $\alpha \to 0$. Another is if the integral is multiplied by the first term in the expansion of
\begin{align}
  (1-z)^{-1+\alpha} = \frac{1}{\alpha} \, \delta(1-z) + \frac{1}{(1-z)_+} + \alpha \left[ \frac{\ln(1-z)}{1-z} \right]_+ + \mathcal{O}(\alpha^2) \, .
\end{align}
In the latter case the expansion of the integral at $z=1$ is needed to $\alpha^1$.
It proves useful to distinguish the two cases $a_{8,9}=0$ and $a_{8,9} \neq 0$. For all integrals with $a_{8,9} \neq 0$ that we encountered, neither of the two conditions above apply, and therefore we can always drop the $\alpha$ regulator in them.
For integrals with $a_{8,9}=0$, we can use the freedom of parameterizing $n$ and $\nb$ to exchange $a_1 \leftrightarrow a_2$ and $a_3 \leftrightarrow a_4$, and always bring the $\alpha$ dependence to $a_1$ and $a_3$. For both cases, we can then use a partial fraction decomposition and the symmetry between $l$ and $k-l$ to reduce $a_4$ to 0.
In the end, what we need to calculate are: $I^{\text{RR}}(a_1,a_2,a_3,0,a_5,a_6,a_7,0,0)$ with $\alpha$ in $a_1$, $a_3$ and possibly also in $a_5$ as well as $I^{\text{RR}}(a_1,a_2,a_3,0,a_5,a_6,a_7,a_8,0)$ and $I^{\text{RR}}(a_1,a_2,a_3,0,a_5,a_6,a_7,0,a_9)$ without $\alpha$ regulator.
\\
The corresponding calculations are further outlined in appendix~\ref{sec:IRR}.
The solutions to most of the relevant integrals can then be obtained straightforwardly, while the remaining solutions are listed in appendix~\ref{sec:list_IRR}.

\section{Results}
\label{sec:results}
Combining the contributions to the NNLO result~\eqref{eq:sB2_sum} expanded up to the finite terms in the regulators $\alpha$ and $\epsilon$, carrying out the refactorization \eqref{eq:refactorization}, identifying the matching kernels \eqref{eq:match_reg} and renormalizing them and the anomaly coefficients (\ref{eq:renorm_I},\,\ref{eq:renorm_F}), we obtain the final results.
In the FORM module \cite{Module} associated with this article, we provide the full set of results in digital form.
Here, we present only the parts which are free of scale logarithms and obtained for $\mu=\mu_x\equiv \frac{2e^{-\gamma_E}}{x_T}$.
These are $F^{(2)}_{i\ib}(L_\perp=0)$ and $I_{i/j}^{(2)}(z,L_\perp=0)$.
The corresponding expressions at $\mu \neq \mu_x$, containing powers of $L_\perp$, can straightforwardly be obtained from these expressions as explained in section~\ref{sec:scale_logs}. 

The NNLO anomaly coefficients result in accordance to \cite{Becher:2010tm} into
\begin{align}
\frac{F^{(2)}_{q\bar{q}}(0)}{C_F} =
\frac{F^{(2)}_{gg}(0)}{C_A} =
&\,
C_A \bigg[ \frac{808}{27} -28\zeta_3 \bigg]
-T_F N_f \frac{224}{27}
\,\raisebox{-5pt}{.}
\label{eq:F2_result}
\end{align}
The NNLO matching kernels are expressed in terms of harmonic polylogarithms $H_{\vec{a}_n} \equiv H_{\vec{a}_n}(z)$ introduced in \cite{Remiddi:1999ew}, $\zeta$ values and functions $\tilde{p}_{ij}$ related to the lowest order DGLAP splitting kernels $P_{ij}^{(0)}$ by eqs.~(\ref{eq:splittingFcts},\,\ref{eq:splittKernels_stripped}).

The gluon-to-gluon kernel is given by
{\small 
\allowdisplaybreaks[1]
\begin{align}
\nonumber
&I_{g/g}^{(2)}(z,0) =
C_A^2
\bigg\{
\delta(1-z)\bigg[
\frac{25}{4} \zeta_{4} -\frac{77}{9} \zeta_{3} -\frac{67}{6} \zeta_{2} +\frac{1214}{81}\bigg]
+
\tilde{p}_{gg}(z)\bigg[
  -4 H_{0,0,0} +8 H_{0,1,0} +8 H_{0,1,1} 
\\
&
 -8 H_{1,0,0} +8 H_{1,0,1} +8 H_{1,1,0} +52 \zeta_{3} -\frac{808}{27}\bigg]
+
\tilde{p}_{gg}(-z)\bigg[
  -16 H_{ -1, -1,0} +8 H_{ -1,0,0} +16 H_{0, -1,0}
\nonumber\\
&
 -4 H_{0,0,0} -8 H_{0,1,0} -8 H_{ -1} \zeta_{2} +4 \zeta_{3}\bigg]
+
\bigg[
 -16 (1 +z) H_{0,0,0} + \frac{8 (1 -z) (11 -z +11 z^2)}{3 z} \big( H_{1,0} +\zeta_{2} \big) 
\nonumber\\
&
 +\frac{2 (25 -11 z +44 z^2)}{3} H_{0,0} -\frac{2 z}{3} H_{1} -\frac{(701 +149 z +536 z^2)}{9} H_{0} +\frac{4 ( -196 +174 z -186 z^2 +211 z^3)}{9 z}   \bigg]
\bigg\}
  \nonumber\displaybreak[2]\\
&
+C_AT_F N_f
\bigg\{
\delta(1-z)\bigg[
\frac{28}{9} \zeta_{3} 
+\frac{10}{3} \zeta_{2} -\frac{328}{81}\bigg]
+
\frac{224}{27}\tilde{p}_{gg}(z)
+
\bigg[
\frac{8 (1 +z)}{3} H_{0,0} +\frac{4 z}{3} H_{1} +\frac{4 (13 +10 z)}{9} H_{0}
\nonumber\\
&
-\frac{4 ( -65 +54 z -54 z^2 +83 z^3)}{27 z}  \bigg]
\bigg\}
\nonumber\displaybreak[2]\\
&
+C_FT_F N_f
\bigg\{
 8 (1 +z) H_{0,0,0} +4 (3 +z) H_{0,0} +24 (1 +z) H_{0} -\frac{8 (1 -z) (1 -23 z +z^2)}{3 z} 
\bigg\}
\nonumber
\displaybreak[3]\,\raisebox{-5pt}{.}
\\
\intertext{\normalsize
The quark-to-gluon kernel reads
}
\nonumber
&I_{g/q}^{(2)}(z,0) =
C_F C_A
\bigg\{
\tilde{p}_{gq}(z)\bigg[
 4 H_{1,1,1}  +4 H_{0,1,1}  +4 H_{1,0,1} +4 H_{1,1,0} +8 H_{0,1,0}
-4 H_{1,0,0} +\frac{44}{3} \big( H_{1,0} +\zeta_{2} \big) 
\nonumber\\
&
-\frac{22}{3} H_{1,1} 
 +\frac{152}{9} H_{1} +24 \zeta_{3} -\frac{1580}{27}
 \bigg]
+
\tilde{p}_{gq}(-z)\bigg[
  -8 H_{ -1, -1,0} +4 H_{ -1,0,0} +8 H_{0, -1,0} -4 H_{ -1} \zeta_{2}\bigg]
\nonumber\\
&
+\bigg[
-4 (2 +z) H_{0,0,0} 
+16 H_{0,1,0} +4 z H_{ -1,0} +4 z H_{0,1} +4 z H_{1,1}   -\frac{8 (1 +z +2 z^2)}{3} H_{1,0}
-\frac{22 z}{3} H_{1}
\nonumber\\
&
 +\frac{2 (36 +9 z +8 z^2)}{3} H_{0,0}  
-\frac{2 (249 -6 z +88 z^2)}{9} H_{0}     
  -8 \zeta_{3} -\frac{2 (4 +13 z +8 z^2)}{3} \zeta_{2}  +\frac{4 (1 +127 z +152 z^2)}{27}\bigg]
\bigg\}
\nonumber\displaybreak[2]\\
&
+C_F^2
\bigg\{
\tilde{p}_{gq}(z)\bigg[
   -4 H_{1,1,1} 
+6 H_{1,1} -16 H_{1}\bigg]
+
\bigg[
2 ( 2 -z) H_{0,0,0} -(4 +3 z) H_{0,0}  -4 z H_{1,1} +6 z H_{1}
\nonumber\\
&
 -5 ( 3 -z) H_{0} 
+( 10 -z) \bigg]
\bigg\}
+C_FT_F N_f
\bigg\{
\tilde{p}_{gq}(z)\bigg[
\frac{8}{3} H_{1,1} -\frac{40}{9} H_{1} +\frac{224}{27}\bigg]
+
\bigg[
\frac{8 z}{3} H_{1} -\frac{40 z}{9}\bigg]
\bigg\}
\nonumber
\displaybreak[3]\,\raisebox{-5pt}{,}
\\
\intertext{\normalsize
while the gluon-to-quark kernel is obtained as
}
\nonumber
&I_{q/g}^{(2)}(z,0) =
C_AT_F
\bigg\{
\tilde{p}_{qg}(z)\bigg[
4 H_{1,0,1}  +4 H_{1,1,0} -4 H_{1,1,1} 
+4 H_{1,1} -\frac{44}{3} H_{0,0} 
+\frac{44}{3} \big( H_{1,0} +\zeta_{2} \big)  +\frac{136}{9} H_{0}
\nonumber\\
&
 +4 H_{1}  -\frac{298}{27}
\bigg]
+
\tilde{p}_{qg}(-z)\bigg[
 -8 H_{ -1, -1,0} +4 H_{ -1,0,0} +8 H_{0, -1,0} +4 H_{ -1,0} -4 H_{ -1} \zeta_{2} \bigg]
\nonumber\\
&
+
\bigg[
4 (1 +2 z) H_{0,0,0} -16 z H_{0,1,0}
 +\frac{2 ( 19 -32 z)}{3} H_{0,0} -4 H_{ -1,0} -4 H_{1,1}
 -\frac{4 (4 +5 z +2 z^2)}{3 z} \big( H_{1,0} +\zeta_{2} \big)
\nonumber\\
&
 +2 ( -2 +z) H_{1} 
-\frac{4 ( 13 -38 z)}{9} H_{0} 
 +8 z (\zeta_{3} +\zeta_{2})  +\frac{2 (172 -166 z +89 z^2)}{27 z} \bigg]
\bigg\}
\nonumber\displaybreak[2]\\
&
+C_FT_F
\bigg\{
\tilde{p}_{qg}(z)\bigg[
4 H_{1,1,1} -4 H_{1,0,0}  +4 H_{0,1,1}  -4 H_{0,0,0}
-4 H_{1,1} -4 H_{1,0} 
 -4 H_{0,1}  -4 H_{0,0}
 -4 H_{1}
\nonumber\\
&
 -4 H_{0}  +28 \zeta_{3} +6 \zeta_{2}  -36\bigg]
+
\bigg[
2 ( 1 -2 z) H_{0,0,0} +(5 +4 z) H_{0,0}
 +4 H_{0,1} +4 H_{1,0} +4 H_{1,1}
 \nonumber\\
&
+2 ( 2 -z) H_{1}  +(12 +7 z) H_{0}
  -6 \zeta_{2} +(23 +3 z) \bigg]
\bigg\}
\nonumber
\displaybreak[3]\,\raisebox{-5pt}{.}
\\
\intertext{\normalsize
The matching kernel of a quark evolving to a quark of the same flavor is given by
}
\nonumber
&I_{q/q}^{(2)}(z,0) =
C_F C_A
\bigg\{
\delta(1-z)\bigg[
5 \zeta_{4} -\frac{77}{9} \zeta_{3} -\frac{67}{6} \zeta_{2}  +\frac{1214}{81} \bigg]
+\tilde{p}_{qq}(z)\bigg[
-2 H_{0,0,0} -4 H_{0,1,0} -4 H_{1,0,1} 
\nonumber\\
&
-4 H_{1,1,0}
-\frac{11}{3} H_{0,0} -\frac{76}{9} H_{0}  +2 \zeta_{3} -\frac{404}{27}\bigg]
+\bigg[
-4 (1 -z) H_{1,0} -4 z H_{0,0} -2 z H_{1} +2 (1 +5 z) H_{0}
\nonumber\\
&
-6 (1 -z) \zeta_{2} +\frac{44}{3} (1 -z) \bigg]
\bigg\}
+C_F^2
\bigg\{
\frac{5}{4} \zeta_{4}\, \delta(1-z)
+\tilde{p}_{qq}(z)\bigg[
8 H_{0,1,0} +4 H_{0,1,1} -4 H_{1,0,0} +8 H_{1,0,1} 
\nonumber\\
&
+8 H_{1,1,0}+3 H_{0,0} +8 H_{0} +24 \zeta_{3}\bigg]
+\bigg[
2 (1 +z) H_{0,0,0} +(3 +7 z) H_{0,0}
 +4 (1 -z) H_{0,1} 
\nonumber\\
&
+12 (1 -z) H_{1,0} 
+2 z H_{1}
 +2 ( 1 -12 z) H_{0}  +6 (1 -z) \zeta_{2}   -22 (1 -z) \bigg]
\bigg\}
\nonumber\displaybreak[2]\\
&
+C_FT_F N_f
\bigg\{
\delta(1-z)\bigg[
\frac{28}{9} \zeta_{3} +\frac{10}{3} \zeta_{2} -\frac{328}{81}\bigg]
+\tilde{p}_{qq}(z)\bigg[
\frac{4}{3} H_{0,0} +\frac{20}{9} H_{0} +\frac{112}{27}\bigg]
-
 \frac{4}{3}(1 -z)
\bigg\}
\nonumber
\\
&
+I_{q'/q}^{(2)}(z,0) \displaybreak[3]\,. \nonumber
\\
\intertext{\normalsize
For a quark evolving to a quark (or anti-quark) of different flavor, it reads instead
}
&I_{q'/q}^{(2)}(z,0) =
C_FT_F
\bigg\{
4 (1 +z) H_{0,0,0} -\frac{2 (3 +3 z +8 z^2)}{3} H_{0,0}
 -\frac{8 (1 -z) (2 -z +2 z^2)}{3 z} \big( H_{1,0} + \zeta_{2} \big)  
\nonumber\\
&
  +\frac{4 (21 -30 z +32 z^2)}{9} H_{0} +\frac{2 (1 -z) (172 -143 z +136 z^2)}{27 z}
\bigg\}
\displaybreak[3]\,\raisebox{-5pt}{,}
\nonumber
\\
\intertext{\normalsize
while for a quark evolving to an anti-quark of the same flavor it is obtained as
}
\nonumber
&I_{\bar{q}/q}^{(2)}(z,0) =
\big(C_F C_A -2C_F^2 \big)
\bigg\{
\tilde{p}_{qq}(-z)\bigg[
8 H_{ -1, -1,0} -4 H_{ -1,0,0} 
-8 H_{0, -1,0}
+4 H_{0,1,0}
+2 H_{0,0,0} 
\nonumber\\
&
+4 H_{ -1} \zeta_{2} -2 \zeta_{3}
\bigg]
+
\bigg[
4 (1 -z) H_{1,0} +4 (1 +z) H_{ -1,0} -(3 +11 z) H_{0}
 +2 ( 3 -z) \zeta_{2} -15 (1 -z) \bigg]
\bigg\}
\nonumber\\
&
+I_{q'/q}^{(2)}(z,0)
\nonumber
\displaybreak[3]\,.
\end{align}
}
In a slightly different notation, we reported these results already in \cite{Gehrmann:2012ze,Gehrmann:2014uaa}.
All other splitting kernels $I_{i/j}^{(2)}$ are related by charge conjugation or flavor symmetry to these results.
The charge conjugation symmetry implies the equality $I_{\bar{\imath}/\bar{\jmath}} = I_{i/j}$ and to respect the flavor symmetry we introduced above only a quark $q$ of unspecified flavor and a quark $q'$ of different flavor.
Moreover, the relation $I_{\bar{q}'/q}=I_{q'/q}$ holds up to NNLO. As a check of our results, we also considered other combinations of partons and found agreement.

\subsection[Relation to $q_T$-resummation in the Collins-Soper framework]{Relation to {\boldmath $q_T$}-resummation in the Collins-Soper framework}
In \cite{Catani:2011kr, Catani:2012qa} the hard-collinear coefficient functions for Drell-Yan and Higgs production were calculated within the framework established in~\cite{Catani:2000vq,Bozzi:2005wk} up to NNLO+NNLL.
A process-independent formulation of this framework for $q_T$-resummation is derived in detail in \cite{Catani:2013tia}. 
The same framework is also used as construction principle for a subtraction scheme~\cite{Catani:2007vq} for fixed-order NNLO calculations. 

Our results are obtained in a completely different approach to $q_T$-resummation, based on a different factorization into individual contributions.
Consequently, the building blocks of the resummed cross section can not be compared one-by-one between the approaches, since they are scheme-dependent.
Both approaches must agree on the scheme-independent expression for the resummed cross section, as we will verify explicitly below. 

In eq.~(6) of \cite{Catani:2013tia}, the differential cross section is expressed in a factorized and resummed form, which contains the hard factor $[ H^F C_1 C_2 ]$.
For $q\bar{q}$ initiated processes the latter is given by the product
\begin{align}
\left[ H^F C_1 C_2 \right]_{q{\bar q};a_1a_2}
 = H_q^F  C_{q a_1}(z_1)   C_{{\bar q} a_2}(z_2) \;,
\end{align}
for $gg$ initiated processes it is given by the following contraction of tensors \cite{Catani:2010pd}
\begin{align}
\left[ H^F C_1 C_2 \right]_{gg;a_1a_2}
&= H_{g,\,\mu_1 \nu_1 \mu_2 \nu_2 }^F
 C_{g a_1}^{\mu_1 \nu_1}(z_1) 
 C_{g a_2}^{\mu_2 \nu_2}(z_2)
\;,
\end{align}
where the dependence on Laplace-space variables and coupling constants has been omitted for clarity. 

In our language, $H^F$ corresponds to the square of the Wilson coefficient $C_F$, which arises on matching QCD on the effective field theory. 
The process-independent factors $C_1$ and $C_2$ correspond to the collinear and anti-collinear matching kernels $I$, respectively.
However, there is no one-to-one correspondence, since these expressions are scheme-dependent. 

Nevertheless, their product  related to the physical cross section by eq.~(6) of \cite{Catani:2013tia} and our eqs.~(\ref{eq:dsigma_DY_full}, \ref{eq:Cqqb_DY}) is well defined after carrying out the convolution in the momentum fractions $z_1$ and $z_2$.
In \cite{Catani:2011kr, Catani:2012qa} this is given  by
\begin{align}
 \mathcal{H}_{ab\leftarrow jk}^F(z,\as ) = \int_0^1 \!\!d z_1 \int_0^1 \!\!d z_2 \, \delta(z-z_1 z_2)  \left[ H^F C_1 C_2 \right]
 \,,
\end{align}
for Drell-Yan and Higgs production respectively. From the process-dependent Wilson coefficients and our results on the process-independent matching kernels, we can determine these $\cal{H}$ functions as 
\begin{align}
  \mathcal{H}^{DY}_{q\bar{q}\leftarrow j k}(z,\as) 
  &= 
  \big|C_V(-q^2,\sqrt{q^2})\big|^2 
  I_{q/j}(z,x_T^2,\mu_x) \otimes I_{\bar{q}/k}(z,x_T^2,\mu_x) \, ,
  \\
 \mathcal{H}^{H}_{gg\leftarrow j k}\big(z,\as,{\textstyle\log\frac{m_t^2}{m_h^2}}\big) 
 &=  
 H^H_{\mu_1\nu_1,\,\mu_2\nu_2}(m_t^2,m_h^2,m_h) \, 
 I_{g/j}^{\mu_1\nu_1}(z,\xp,\mu_x) \otimes I_{g/k}^{\mu_2\nu_2}(z,\xp,\mu_x)  \, ,
 \label{eq:H2_H}
 \end{align}
where each function is evaluated at a value of the renormalization scale for which no large logarithms arise, which is the invariant mass of the produced final state and $\mu_x = 2 e^{-\gamma_E}/x_T$, respectively.
In the second line, $I^{\mu\nu}_{g/j}$ is the gluon matching tensor which is related to $\Bcal_{g/N}^{\mu\nu}$ in \eqn{eq:Bcal_gluon} in a completely analogous way as $I_{g/j}$ is related to $\Bcal_{g/N}$ \cite{Becher:2012yn,Gehrmann:2014uaa}.
It can be decomposed into the two independent components $I_{g/j}$ and $I'_{g/j}$ analogously to \eqn{eq:tensor_compo}. 
$H_H^{\mu_1\nu_1,\,\mu_2\nu_2}$ is the hard tensor. For Higgs production it has the explicit form
\begin{align}
 H^H_{\mu_1\nu_1,\,\mu_2\nu_2}(m_t^2,m_h^2,m_h)
 =
 C_t^2(m_t^2,m_h) \big|C_S(-m_h^2,m_h)\big|^2
 g_{\mu_1 \mu_2} g_{\nu_1 \nu_2}
 \,,
\end{align}
with the Wilson coefficients arising on first integrating out the top quark and then matching to SCET.
To determine \eqn{eq:H2_H} to NNLO, the NLO results of $I'_{g/j}$ are required which we calculated finding results in accordance with \cite{Becher:2012yn}.

The resulting expressions for the $\mathcal{H}$ coefficients are found in full agreement with the results in  \cite{Catani:2011kr, Catani:2012qa}, and constitute a fully independent validation of them in a completely different calculational approach. 

\subsection{Further checks}
Below we describe further observations and checks confirming our results for the matching kernels $I^{(n)}$ and anomaly coefficients $F^{(n)}$ with $n\leq 2\,$.
We first observe that these functions depend only on the scale logarithm $L_\perp$ and the momentum fraction $z$.
As required by consistency, no dependence on the analytic regulator $\alpha$ or the associated scale $\nu$ remained, but they canceled in \eqn{eq:refactorization}, where moreover all dependence on the hard scale $q^2$ had been refactorized from the resulting functions.
This not only confirms our results but also the consistency of the whole framework and explicitly demonstrates the applicability of the analytic regulator of \cite{Becher:2011dz} in high order calculations.

Moreover, in our results no poles in the dimensional regulator $\epsilon$ remained, but they could consistently be removed by renormalization (\ref{eq:renorm_F},\,\ref{eq:renorm_I}), where the exact renormalization factors had been implied already by their RGEs in terms of known functions and are listed in section~\ref{sec:ren_factors}.
We also explicitly confirmed that $F_{i\ib}(\lp,\as)$ and $I_{i/j}(z,\lp,\as)$ themselves obey the RGEs~(\ref{eq:RGE_F},\,\ref{eq:RGE_I}) and that their $\lp$ dependent terms can be reconstructed through the relations in appendix~\ref{sec:scale_logs} from the results listed here and the expressions in appendix~\ref{sec:ad} and \ref{sec:lower_order}.
These points are yet another strong confirmation of our results.

Furthermore, we did not only perform the calculation in light cone gauge as described in this article, but also in Feynman gauge finding identical results.
This not only serves as test to our calculation, but also explicitly demonstrates that the individual factors in our framework are gauge invariant.

In addition to that, we compared our results to literature: we could explicitly confirm the expressions for the anomaly coefficients and the NLO matching kernels as given in \cite{Becher:2010tm, Becher:2012yn}.
\section{Conclusions}
\label{sec:conclusions}
In this paper, we have derived perturbative QCD corrections to all parton-to-parton TPDFs at NNLO. Our calculation is based on a gauge invariant operator definition \cite{Becher:2010tm, Becher:2012yn} with an analytic regulator \cite{Becher:2011dz}.
We demonstrate for the first time that such a definition works beyond the first non-trivial order, and that it provides a fully complementary approach to $q_T$-resummation in the CSS framework~\cite{Collins:1984kg, Catani:2000vq,  Bozzi:2005wk, Catani:2013tia}. From our calculation, we extract the coefficient functions relevant for $q_T$-resummation at N$^3$LL accuracy. 
Our results can be applied to any process yielding a colorless final state, provided the NNLO virtual corrections are known.
They confirm the recent structural findings in~\cite{Catani:2013tia}, while working with a completely different methodology \cite{Becher:2010tm, Becher:2011dz, Becher:2012yn} based on SCET. 
Combined with the work of \cite{Zhu:2012ts}, our results could also be applied to the transverse momentum resummation in  $t\bar{t}$ production.
For gluon-gluon initiated processes with a general spin structure, in addition to the results presented here, N$^3$LL transverse momentum resummation may require the NNLO corrections to the second tensor structure of the gluon TPDFs, which we will present in a separate article. 
We documented our calculation in detail, and validated our results with numerous non-trivial checks, including an independent re-derivation of the second-order contributions to the hard factors ${\cal H}$ for Drell-Yan and Higgs production that were obtained previously in~\cite{Catani:2011kr, Catani:2012qa}. 
A digital form of our results is provided in \cite{Module}.

\acknowledgments

We would like to thank Matthias Neubert, Guido Bell and  Massimiliano Grazzini for useful discussions.
T.L.\ would like to thank Peking University for the hospitality during part of this work has been completed.
This work was supported in part by the Schweizer Nationalfonds under grant 200020-141360/1, by the Research Executive Agency (REA) of the European Union under the Grant Agreement number PITN-GA-2010-264564 (LHCPhenoNet), by the National Natural Science Foundation of China under Grant No.\ 11345001, and by the Bundesministerium f\"ur Bildung und Forschung through contract (05H12GU8).

\appendix

\section{Calculation of double real integrals}
\label{sec:IRR}
In this appendix we outline the determination of the integrals $I^{\text{RR}}(\{a_i\})$ defined in \eqn{eq:IRR} which appear for the double real emission. 
As explained in  section~\ref{sec:2R}, we distinguish three relevant subsets of integrals.

We first consider the integrals with $a_{8,9}=0$. It is convenient to define integrals of the form
\begin{align}
  \label{eq:IRR1IRR2}
  I_1^{\text{RR}}(a_1,a_2,a_3) =& \frac{1}{2\pi} \, \frac{\Gamma^2(1-\epsilon)}{\Gamma(1-2\epsilon)} \int_0^\pi \!\!d\theta_1 \sin^{1-2\epsilon}\!\theta_1 \int_0^\pi \!\!d\theta_2 \sin^{-2\epsilon}\!\theta_2  \, D_1^{-a_1} \, D_2^{-a_2} \, D_3^{-a_3} \, , \nonumber
\displaybreak[1]
  \\
  I_2^{\text{RR}}(a_5,a_6,a_7) \equiv& \int_0^1 \!\!d y \, D_5^{-a_5+\epsilon} \, D_6^{-a_6-\epsilon} \, D_7^{-a_7} 
= z^{1-a_5+\epsilon-a_7} \, \frac{\Gamma(1-a_5+\epsilon)\Gamma(1-a_6-\epsilon)}{\Gamma(2-a_5-a_6)}
\nonumber
\\ 
& \times {}_2F_1 \left( 1-a_5+\epsilon, 2-a_5-a_6-a_7; 2-a_5-a_6; 1-z \right) .
\end{align}
The full integrals are then given by
\begin{align}
  \label{eq:IRRIRR1}
  I^{\text{RR}}(a_1,a_2,a_3,0,a_5,a_6,a_7,0,0) &= \int_0^1 \!\!d y \, I_1^{\text{RR}}(a_1,a_2,a_3) \, D_5^{-a_5+\epsilon} \, D_6^{-a_6-\epsilon} \, D_7^{-a_7} \, .
\end{align}
If one of its arguments is 0, the $I_1^{\text{RR}}$ integrals can be readily calculated to be
\begin{align}
  I_1^{\text{RR}}(a_1,a_2,0) &= \frac{\Gamma(1-\epsilon-a_1) \, \Gamma(1-\epsilon-a_2)}{\Gamma(2-2\epsilon-a_1-a_2)} \, {}_2F_1 \left( a_1, a_2; 1-\epsilon; y \right) , \nonumber
    \displaybreak[1]
  \\
  I_1^{\text{RR}}(0,a_2,a_3) &= \frac{\Gamma(1-\epsilon-a_2) \, \Gamma(1-\epsilon-a_3)}{\Gamma(2-2\epsilon-a_2-a_3)} \, {}_2F_1 \left( a_3, a_2; 1-\epsilon; 1-y \right) , \nonumber
    \displaybreak[1]
  \\
  I_1^{\text{RR}}(a_1,0,a_3) &= \frac{\Gamma(1-\epsilon-a_1) \, \Gamma(1-\epsilon-a_3)}{\Gamma(2-2\epsilon-a_1-a_3)} \, .
\end{align}
If furthermore $a_7=0$, the remaining integral over $y$ can be carried out, and the result is
\begin{align}
  I^{\text{RR}}(a_1,a_2,0,0,a_5,a_6,0,0,0) = &\,
 \frac{\Gamma(1-\epsilon-a_1) \, \Gamma(1-\epsilon-a_2) \, \Gamma(1-a_5+\epsilon) \, \Gamma(1-a_6-\epsilon)}{\Gamma(2-2\epsilon-a_1-a_2) \, \Gamma(2-a_5-a_6)} 
\nonumber\\ 
&\times{}_3F_2 \left( a_1,a_2,1-a_5+\epsilon; 1-\epsilon,2-a_5-a_6; 1 \right) , 
  \nonumber\\
  I^{\text{RR}}(0,a_2,a_3,0,a_5,a_6,0,0,0) = &\, \frac{\Gamma(1-\epsilon-a_2) \, \Gamma(1-\epsilon-a_3) \, \Gamma(1-a_5+\epsilon) \, \Gamma(1-a_6-\epsilon)}{\Gamma(2-2\epsilon-a_2-a_3) \, \Gamma(2-a_5-a_6)} 
\nonumber\\ 
&\times {}_3F_2 \left( a_2,a_3,1-a_6-\epsilon; 1-\epsilon,2-a_5-a_6; 1 \right) .
\label{eq:IRR3F2}
\end{align}
We also have
\begin{align}
  \label{eq:IRRa20}
  I^{\text{RR}}(a_1,0,a_3,0,a_5,a_6,a_7,0,0) = I_1^{\text{RR}}(a_1,0,a_3) \, I_2^{\text{RR}}(a_5,a_6,a_7) \, .
\end{align}
For more generic cases, we change variables to
\begin{align}
  u = \frac{1+\cos\theta_1}{2} \, , \quad v = \frac{1+\cos\theta_2}{2} \, ,
\end{align}
which allows us to rewrite the integral as
\begin{align}
  \label{eq:IRR1}
  I_1^{\text{RR}}(a_1,a_2,a_3) &= \frac{2^{-4\epsilon}}{\pi} \, \frac{\Gamma^2(1-\epsilon)}{\Gamma(1-2\epsilon)} \int_0^1 \!\!d u \!\int_0^1\!\! d v \, u^{-\epsilon-a_3} \, (1-u)^{-\epsilon-a_1} \, v^{-1/2-\epsilon} \, (1-v)^{-1/2-\epsilon} \nonumber
  \\
  &\hspace{1em} \times \left[ \left( \sqrt{u(1-y)} - \sqrt{y(1-u)} \right)^2 + 4v \sqrt{u(1-u)y(1-y)} \right]^{-a2} \, .
\end{align}
>From this representation, it is obvious that if $a_2 \leq 0$, the integrand can be expanded and written in terms of powers of $u$, $1-u$, $v$, $1-v$, $y$ and $1-y$. The integrals over $u$ and $v$ then lead to some $\Gamma$ functions, while the powers of $y$ and $1-y$ can be absorbed into $a_5$ and $a_6$.
The remaining $y$ integral can then be performed with the help of \eqn{eq:IRR1IRR2}.
\\
For $a_2 > 0$, we first perform the $v$ integral to get
\begin{align}
  I_1^{\text{RR}}(a_1,a_2,a_3) &= y^{-a_2} \int_0^y \!d u \, u^{-\epsilon-a_3} \, (1-u)^{-\epsilon-a_1-a_2} \, {}_2F_1 \left( a_2, a_2+\epsilon; 1-\epsilon; {\textstyle \frac{u(1-y)}{y(1-u)} }\right) \nonumber
  \\
  &\hspace{-5em} + (1-y)^{-a_2} \int_y^1 \!d u \, u^{-\epsilon-a_3-a_2} \, (1-u)^{-\epsilon-a_1} \,  {}_2F_1 \left( a_2, a_2+\epsilon; 1-\epsilon; {\textstyle \frac{y(1-u)}{u(1-y)} } \right) .
\end{align}
For each of the two integrals above, we change variable from $u$ to the last argument of the hypergeometric function, which we call $t$. 
We then apply ${}_2F_1(a,b;c;t) = (1-t)^{c-a-b}  {}_2F_1(c-a,c-b;c;t)$ and insert the resulting expression into \eqn{eq:IRRIRR1} to arrive at
\begin{align}
  I^{\text{RR}}(a_1,a_2,a_3,0,a_5,a_6,a_7,0,0) &= \int_0^1 \!\!d y \!\int_0^1\!\! d t \, y^{1-a_2-a_3-a_5} \, (1-y)^{1-2\epsilon-a_1-a_2-a_6} 
  \\
  &\hspace{-12em} \times D_7^{-a_7} \, (1-t)^{1-2a_2-2\epsilon} \, {}_2F_1(1-a_2-\epsilon, 1-a_2-2\epsilon; 1-\epsilon; t)
  \nonumber\\
  &\hspace{-12em} \times \left\{ t^{-\epsilon-a_3} \left[ 1-(1-t)y \right]^{-2+2\epsilon+a_1+a_2+a_3} + t^{-\epsilon-a_1} \left[1-(1-t)(1-y) \right]^{-2+2\epsilon+a_1+a_2+a_3} \right\} .
  \nonumber
\end{align}
From here, the remaining integrals in general cannot be performed in closed form, and a series expansion in $\alpha$ and $\epsilon$ is required. These expansions are documented  in the next appendix.
\medskip

We now turn to the cases where $a_8>0$ or $a_9>0$. As mentioned above, we can always drop the analytic regulator $\alpha$ for these integrals. Therefore we can always reduce $a_4$ to 0.
Following the same procedure as before, cases with $a_2 \leq 0$ can be performed straightforwardly.
For $a_2,\,a_8>0$ we obtain
\begin{align}
  &I^{\text{RR}}(a_1,a_2,a_3,0,a_5,a_6,a_7,a_8,0) = \int_0^1 \!\!d u \!\int_0^1\!\! d y \, y^{-a_5+\epsilon} \, (1-y)^{-a_6-\epsilon} \, D_7^{-a_7} \, D_8^{-a_8} \nonumber
  \\
  &\quad \times \bigg[ \theta(y-u) \, y^{-a_2} \, u^{-a_3-\epsilon} \, (1-u)^{-a_1-a_2-\epsilon} \, {}_2F_1 \left( a_2, a_2+\epsilon; 1-\epsilon; {\textstyle \frac{u(1-y)}{y(1-u)} }\right) \nonumber
  \\
  &\quad\quad + \theta(u-y) \, (1-y)^{-a_2} \, u^{-a_2-a_3-\epsilon} \, (1-u)^{-a_1-\epsilon} \, {}_2F_1 \left( a_2, a_2+\epsilon; 1-\epsilon; {\textstyle   \frac{y(1-u)}{u(1-y)} }\right) \bigg] \, .
\end{align}
The main complication here is that $D_7=[1-(1-y)(1-z)]$ and $D_8=[1-(1-u)(1-z)]$ may both appear. This prevents us from changing variable to the last argument of the hypergeometric function, since regardless of whether we substitute $u$ or $y$, either $D_7$ or $D_8$ will become very complicated in terms of the new variable $t$. 
We therefore now consider specific cases. For $a_7=0$ one obtains
\begin{align}
  &I^{\text{RR}}(a_1,a_2,a_3,0,a_5,a_6,0,a_8,0) = \int_0^1 \!\!d u \!\int_0^1\!\! d t \, u^{1-a_2-a_3-a_5} \, (1-u)^{1-a_1-a_2-a_6-2\epsilon} \, D_8^{-a_8} \nonumber
  \\
  & \times (1-t)^{1-2a_2-2\epsilon} \, {}_2F_1(1-a_2-\epsilon,1-a_2-2\epsilon;1-\epsilon;t) \nonumber
  \\
  & \times \left\{ t^{-a_6-\epsilon} \left[ 1-(1-t)(1-u) \right]^{-2+a_2+a_5+a_6} + t^{-a_5+\epsilon} \left[ 1-(1-t)u \right]^{-2+a_2+a_5+a_6} \right\} .
\end{align}
The representation of $I^{\text{RR}}(a_1,a_2,a_3,0,a_5,a_6,0,0,a_9)$ is essentially the same as above, with $D_8^{-a_8}$ replaced by $D_9^{-a_9}$.

The relevant cases, where in addition $a_7>0$, are $a_7=1$, $a_1,a_3=0$ and either $a_8$ or $a_9=1$. We then partial fraction decompose $D_7$ with $D_8$ or $D_9$, respectively. After changing variables from either $u$ or $y$ to the last argument of the hypergeometric function, which we call $t$, and if relevant renaming $y$ to $u$, one obtains
\begin{align}
&I^\mathrm{RR}(0,a_2,0,0,a_5,a_6,1,1,0) 
=
\int_0^1\!\!d u \!\int_0^1\!\!d t
\,u^{-a_2-a_5-\epsilon} (1-u)^{1-a_2-a_6-\epsilon}D_8^{-1}  
\nonumber\\&
\times ( 1-t)^{-2a_2-2\epsilon}{}_2F_1 \left( 1-a_2-\epsilon,1-a_2-2\epsilon;1-\epsilon;t\right)
\nonumber\\&\times
\Big\{ 
t^{-\epsilon}\left[1- (1-t)(1-u)\right]^{-1+a_2+2\epsilon}
+ t^{-a_6}\left[1-(1-t)(1-u)\right]^{-1+a_2+a_5+a_6}
\nonumber\\&\quad
-t^{-\epsilon}\left[1- (1-t)u\right]^{-1+a_2+2\epsilon} 
-t^{-a_5}\left[1- (1-t)u\right]^{-1+a_2+a_5+a_6}
\Big\}
\label{eq:Irr8_a78eq11}
\end{align}
and an even more involved version of this for $a_9=1$.
At intermediate steps, an additional regulator is introduced in $a_5$ which however does not lead to poles in the final result for the integral.

\section{List of double real integrals}
\label{sec:list_IRR}
In the previous appendix we described the methods of calculating the double real integrals. Some integrals can be represented in an exact form in terms of hypergeometric functions ${}_3F_2$ as in \eqn{eq:IRR3F2}. Several other integrals with $a_2<0$ can be obtained following the steps explained below \eqn{eq:IRR1}.
For other integrals, we calculate them as a series expansion, and list them in this appendix. The results will be written in terms of harmonic polylogarithms $H_{\vec{a}_n} \equiv H(\vec{a}_n,z)$ introduced in \cite{Remiddi:1999ew}.

To which order in $\alpha$ and $\epsilon$ a given integral is needed relies on the prefactor multiplying the integral. We first list the integrals which are needed to order $\alpha^1$. 
We found that they all have $a_7,a_8,a_9=0$, and therefore we will suppress these arguments below.
For these integrals, it is more convenient to choose $\alpha/\epsilon$ instead of $\alpha$ as one of the expansion parameters, since we need to send $\alpha$ to 0 before $\epsilon$. The results are
\begin{align*}
  &I^{\text{RR}}(\alpha,1,\alpha,0,r,1) = \frac{1}{\epsilon^2} - 2\epsilon \zeta_{3} - 3\epsilon^2 \zeta_{4} 
  \\
  &\hspace{6em} - \frac{\alpha}{\epsilon} \bigg[ \frac{1}{2\epsilon^2}  + (1+x)\zeta_{2}  + \epsilon(4-x) \zeta_{3} + \epsilon^2\frac{11+2x}{2} \zeta_{4} \bigg] + \mathcal{O}\big([\alpha/\epsilon]^2,\epsilon^3\big) \, ,
 \displaybreak[1]
 \\
  &I^{\text{RR}}(1+\alpha,1,\alpha,0,r,0) = \frac{1}{\epsilon^2} + 2\zeta_{2} + 4\epsilon \zeta_{3} + 11\epsilon^2  \zeta_{4}
  \\
  &\hspace{6em} - \frac{\alpha}{\epsilon} \bigg[ \frac{1}{\epsilon^2}  + 2x\zeta_{2} - 2\epsilon \zeta_{3} - \epsilon^2  \frac{27-17x}{2} \zeta_{4} \bigg] + \mathcal{O}\big([\alpha/\epsilon]^2,\epsilon^3\big) \, ,
\displaybreak[1]
  \\
  &I^{\text{RR}}(\alpha,1,1+\alpha,0,-1+r,1) = \frac{2}{\epsilon^2} - 2\zeta_{2} - 6\epsilon\zeta_{3} - 8\epsilon^2 \zeta_{4}
  \\
  &\hspace{6em} - \frac{\alpha}{\epsilon} \bigg[ \frac{3}{2\epsilon^2} + 2(1+x) \zeta_{2} + \epsilon (11+2x) \zeta_{3} + \epsilon^2 \frac{56-x}{2}\zeta_{4} \bigg] +\mathcal{O}\big([\alpha/\epsilon]^2,\epsilon^3\big) \, ,
\displaybreak[1]
  \\
  &I^{\text{RR}}(\alpha,1,1+\alpha,0,r,0) = \frac{2x}{\alpha/\epsilon} \bigg[ \frac{1}{\epsilon^2} - 2\epsilon  \zeta_{3} - 3\epsilon^2 \zeta_{4} \bigg] - \frac{x}{\epsilon^2} - 2\zeta_{2} + \epsilon(2-6x) \zeta_{3} - \epsilon^2  (2+9x) \zeta_{4}
  \\
  &\hspace{6em} + \frac{\alpha}{\epsilon} \bigg[ \frac{x}{\epsilon^2} + \epsilon(2-4x) \zeta_{3} - \epsilon^2 \frac{15-5x}{2} \zeta_{4} \bigg] +\mathcal{O}\big([\alpha/\epsilon]^2,\epsilon^3\big) \, ,
\end{align*}
where $r=-\alpha(1-x)$ with $x=\pm 1$. Obviously the last integral contains a pole in $\alpha$. 
For all the remaining integrals we can drop the $\alpha$ regulator and the following results are understood up to corrections of $\mathcal{O}(\alpha)$. The remaining integrals with $a_8,a_9=0$ are
\begin{align*}
  I^{\text{RR}}(1,1,0,0,0,-1,1,0,0) &= \frac{2}{1-z} \, \bigg[ \frac{H_{0}}{\epsilon} + H_{0,0} - H_{1,0} - \zeta_{2}
  \\
  &\hspace{3em} + \epsilon  \big( H_{0,0,0} - 2 H_{0,1,0} - H_{1,0,0} - 3 \zeta_{3} \big) \bigg] + \mathcal{O}(\epsilon^2) \, ,
\displaybreak[1]
  \\
  I^{\text{RR}}(0,1,1,0,-1,0,1,0,0) &= \frac{2}{1-z} \, \bigg[ \frac{H_{0}}{\epsilon} - H_{1,0} - \zeta_{2} + \epsilon \big( -H_{1,0,0} + \zeta_{3} \big) \bigg] + \mathcal{O}(\epsilon^2) \, .
\end{align*}
Note that while the above integrals contain an explicit $(1-z)$ in the denominators, this divergence at $z \to 1$ is canceled by the terms in the numerator and the whole integral is at most logarithmically divergent. The remaining integrals with $a_8 > 0$ or $a_9 > 0$ are
\begin{align*}
  I^{\text{RR}}(-1,1,0,0,0,1,0,1,0) &= \frac{1}{1-z} \bigg[ \frac{2H_{0}}{\epsilon} - 4H_{1,0} - 4\zeta_{2} + 8\epsilon \big(H_{1,1,0} + \zeta_{2} H_{1} - \zeta_{3}\big) \bigg] \, ,
  \displaybreak[1]\\
  I^{\text{RR}}(0,1,0,0,0,0,0,1,0) &= \frac{1}{1-z} \bigg[ \frac{H_{0}}{\epsilon} + H_{0,0} - \epsilon \big(H_{0,0,0} + 2H_{0,1,0} + 2\zeta_{2} H_{0} + 4\zeta_{3} \big) \bigg] \, ,
  \displaybreak[1]\\
  I^{\text{RR}}(0,1,0,0,0,0,0,0,1) &= \frac{1}{1-z} \bigg[ \frac{H_{0}}{\epsilon} - H_{0,0} + \epsilon \big(H_{0,0,0} + 2H_{0,1,0} + 2\zeta_{2} H_{0} + 4\zeta_{3}\big) \bigg] \, ,
  \displaybreak[1]\\
  I^{\text{RR}}(0,1,0,0,0,0,1,0,1) &= \frac{1}{1-z^2} \bigg[ \frac{2H_{0}}{\epsilon} - 4H_{-1,0} + 2H_{0,0} - 2\zeta_{2}
  \\
  &\hspace{-8em} + 2\epsilon \big(4H_{-1,-1,0} - 2H_{-1,0,0}- 4H_{0,-1,0} + H_{0,0,0} + 2H_{0,1,0} + 2\zeta_{2} H_{-1} - \zeta_{3}\big) \bigg] \, ,
  \displaybreak[1]\\
  I^{\text{RR}}(0,1,0,0,0,0,1,1,0) &= -\frac{1}{\epsilon z} + \frac{2H_0}{1-z} + \frac{2}{z} + 2\epsilon \left[ \frac{H_{0,0}-2H_0}{1-z} + \frac{H_{1,0}+\zeta_{2}-2}{z} \right] ,
  \displaybreak[1]\\
  I^{\text{RR}}(0,2,0,0,0,0,1,1,0) &= \frac{(1-z)^2}{6\epsilon z^2} +  \frac{3-z}{1-z} \frac{H_{0}}{3} - \frac{z^2+z+10}{9z^2}
  \\
  &\hspace{-11em} + \frac{\epsilon}{3} \bigg[ \frac{3-z}{1-z} H_{0,0} - \frac{(1-z)^2}{z^2} (H_{1,0}+\zeta_{2}) +\left( \frac{2z}{1-z} +\frac{3}{z} \right) \frac{H_{0}}{3} +\frac{2z^2+11z+47}{9z^2} \bigg] \, ,
  \displaybreak[1]\\
  I^{\text{RR}}(-1,2,-1,0,0,0,0,1,0) &= 0 \, ,
  \displaybreak[1]\\
  I^{\text{RR}}(-1,2,-1,0,0,0,0,2,0) &= 0 \, ,
\end{align*}
where the dropped corrections are of $\mathcal{O}(\epsilon^2)$.

\section{Anomalous dimensions and splitting functions}
\label{sec:ad}
In this appendix we collect the expressions for the anomalous dimensions and splitting functions for the reader's convenience.
We define the perturbative expansion of the quark and gluon anomalous dimensions $\gamma^i$ as
\begin{align}
  \gamma^i(\alpha_s) &= \asFPi \, \gamma^i_0 + \left( \asFPi \right)^2 \gamma^i_1 + \sO(\as^3) \, ,
\end{align}
and analogously for the cusp anomalous dimensions $\Gamma^i$ in the fundamental and adjoint representation.
The coefficients up to the second order are given by
\begin{align}
  &\frac{1}{C_f}\Gamma^q_0 = \frac{1}{C_a}\Gamma^g_0 = 4 \, , \nonumber
  \\
  &\frac{1}{C_f}\Gamma^q_1 = \frac{1}{C_a}\Gamma^g_1 = \left( \frac{268}{9} - \frac{4\pi^2}{3} \right) C_a - \frac{80}{9} T_f N_f \, , \nonumber
\displaybreak[1]
  \\
  &\gamma^q_0 = - 3 C_f \, , \nonumber
  \\
  &\gamma^q_1 = C_f^2 \left( -\frac{3}{2} + 2 \pi^2 - 24 \zeta_3 \right) + C_f C_a \left( - \frac{961}{54} - \frac{11\pi^2}{6} + 26 \zeta_3 \right) + C_f T_f N_f \left( \frac{130}{27} + \frac{2\pi^2}{3} \right) , \nonumber
\displaybreak[1]
  \\
  &\gamma^g_0 = - \frac{11}{3} C_a + \frac{4}{3} T_f N_f \, , \nonumber
  \\
  &\gamma^g_1 = C_a^2 \left( - \frac{692}{27} + \frac{11\pi^2}{18} + 2 \zeta_3 \right) + C_a T_f N_f \left( \frac{256}{27} - \frac{2\pi^2}{9} \right) + 4 C_f T_f N_f \, .
\end{align}
The QCD $\beta$ function is
\begin{align}
  \beta(\alpha_s) &= \frac{d \alpha_s(\mu)}{d \log \mu} = - 2\alpha_s \left[ \asFPi \, \beta_0 + \left( \asFPi \right)^2 \beta_1 + \cdots \right] ,
\displaybreak[1]
\intertext{where}
  \beta_0 &= \frac{11}{3} C_A - \frac{4}{3} T_F N_f \, , \nonumber
  \\
  \beta_1 &= \frac{34}{3} C_A^2 - \frac{20}{3} C_A T_F N_f - 4 C_F T_F N_f \, .
\end{align}
Higher order coefficients of $\Gamma^i,\,\gamma_i$ and $\beta$ can be found in \cite{Moch:2004pa,Becher:2009qa,vanRitbergen:1997va}, respectively.
\\
The DGLAP splitting functions are
\begin{align}
  P_{ij}(z,\mu) &= \asFPi \, P_{ij}^{(0)}(z) + \left( \asFPi \right)^2 P_{ij}^{(1)}(z) + \cdots \, ,
\displaybreak[1]
\intertext{
where the first order coefficients are
}
  P_{qq}^{(0)}(z) &= 2C_F\tilde{p}_{qq}(z) + 3C_F \delta(1-z) , \nonumber
  \\
  P_{gg}^{(0)}(z) &= 4C_A  \tilde{p}_{gg}(z) + \left[ \tfrac{11}{3} C_A - \tfrac{4}{3} T_F N_f \right] \delta(1-z) \, , \nonumber
\displaybreak[1]
  \\
  P_{qg}^{(0)}(z) &= 2 T_F \tilde{p}_{qg}(z) \, , \nonumber
  \\
  P_{gq}^{(0)}(z) &= 2 C_F \tilde{p}_{gq}(z) \, , 
  \label{eq:splittingFcts}
\displaybreak[3]
\intertext{
with the functions
}
\tilde{p}_{qq}(z) &=  \frac{1+z^2}{(1-z)_+}
\,\raisebox{-3pt}{,}
\nonumber\\
\tilde{p}_{gg}(z) &= \frac{z}{(1-z)_+} + \frac{1-z}{z} + z(1-z)
\,\raisebox{-3pt}{,}
\nonumber\\
\tilde{p}_{qg}(z) &=  z^2 + (1-z)^2 
\,\raisebox{-3pt}{,}
\nonumber\\
\tilde{p}_{gq}(z) &=  \frac{1+(1-z)^2}{z}
\label{eq:splittKernels_stripped}
\,\raisebox{-3pt}{.}
\end{align}
The second order coefficients can be obtained from the results in \cite{Curci:1980uw, Furmanski:1980cm} and we do not repeat these expressions here. The coefficients up to third order are given in \cite{Moch:2004pa, Vogt:2004mw}.

By $(\ldots)_+$ we denote the plus prescription with support on $[0,1]$ regulating the pole at $z=1$. 
To express our results, we also use $\tilde{p}_{ij}(-z)$. In those cases the plus prescription is dropped.

\section{Further results}
\label{sec:outsourced_results}
In this section, we collect a number of results which either appear in intermediate steps or have been given in the literature already.
Due to the flavor and charge conjugation symmetry of QCD, the number of independent functions reduces and below we use a notation, where $q$ refers to a quark of unspecified (but same) flavor and $q'$ to a quark of different flavor. Up to NNLO we moreover have $\Bcal_{\bar{q}/q}=\Bcal_{\bar{q}'/q}$ and corresponding relations for the other functions.

\subsection{Renormalization factors}
\label{sec:ren_factors}
As explained in section~\ref{sec:framework}, the renormalization factors are related by their RGEs to the anomalous dimensions and QCD $\beta$ function listed in appendix~\ref{sec:ad}.
Here we list the resulting expressions for the perturbative coefficients according to \eqn{eq:pert_expansion} up to NNLO for $\phi_{i/j}$, $Z^B_i$ and $Z^F_i$ 
beyond their LO terms 
\begin{align}
 \phi^{(0)}_{i/j}(z) = \delta_{i/j}\delta(1-z)\,,\;\;\;
 Z^{B,(0)}_i = 1\,,\;\;\;
 Z^{F,(0)}_i = 0\,.
\end{align}
Due to \eqn{eq:phi_bare}, the parton-to-parton PDFs are directly related to their renormalization factors and due to the RGEs both of them to the DGLAP splitting kernels, yielding the relations
\begin{align}
  \phi_{i/j}^{(1)}(z) &= -\frac{P_{ij}^{(0)}(z)}{\epsilon} \, , \nonumber
  \\
  \phi_{i/j}^{(2)}(z) &= \frac{1}{2\epsilon^2} \left[ \sum_k P_{ik}^{(0)}(z) \otimes P_{k j}^{(0)}(z) + \beta_0 \, P_{ij}^{(0)}(z) \right] - \frac{P_{ij}^{(1)}(z)}{2\epsilon} 
  \,\raisebox{-5pt}{.}
\end{align}
For $Z^B_i$ in \eqn{eq:renorm_B} we find
\begin{align}
  Z^{B,(1)}_i(L_\perp) &= \frac{\Gamma^i_0}{2\epsilon^2} + \frac{\Gamma^i_0 \, L_\perp - 2\gamma^i_0}{2\epsilon} \, , \nonumber
  \\
  \label{eq:Z_B}
  Z^{B,(2)}_i(L_\perp) &= \frac{(\Gamma^i_0)^2}{8\epsilon^4} + \frac{\Gamma^i_0}{4\epsilon^3} \left[ \Gamma^i_0 \, L_\perp - 2\gamma^i_0 - \frac{3}{2} \, \beta_0 \right]
  \\
  &+ \frac{1}{8\epsilon^2} \left[ \Gamma^i_1 + \left( \Gamma^i_0 \, L_\perp - 2\gamma^i_0 \right)^2 - 2\beta_0 \left( \Gamma^i_0 \, L_\perp - 2\gamma^i_0 \right) \right] + \frac{1}{4\epsilon} \left( \Gamma^i_1 \, L_\perp - 2\gamma^i_1 \right) , \nonumber
\end{align}
while the coefficients for $Z^F_i$ in \eqn{eq:renorm_F} are given by
\begin{align}
  Z^{F,(1)}_i &= \frac{\Gamma^i_0}{\epsilon} \, , \nonumber
  \\
  Z^{F,(2)}_i &= -\frac{\beta_0\Gamma^i_0}{2\epsilon^2} + \frac{\Gamma^i_1}{2\epsilon} 
 \, .
\end{align}
\subsection{Dependence on scale logarithms}
\label{sec:scale_logs}
The $\lp$ dependence of $F_{i\ib}$ and $I_{i/j}$ can be recovered from their values at $\lp=0$ by solving the RGEs~(\ref{eq:RGE_F},\,\ref{eq:RGE_I}).
More explicitly, we can expand these functions in both $\as$ and $\lp$ according to
\begin{align}
 F_{i\ib}(\lp,\as) &= \sum_{n \geq 1} \sum_{l=0}^{n}  F_{i\ib}^{(n,l)} \left(\asFPi\right)^n \lp^l
 \label{eq:expand_Fi}
\,,
\\
I_{i/j}(z,\lp,\as) &= 
 \sum_{n \geq 0} \sum_{l=0}^{2n}  I^{(n,l)}_{i/j}\!(z)\; \left(\asFPi\right)^n \lp^l
 \label{eq:expand_Iij}
 \,.
\end{align}
Since the RGEs have to hold for all values of $\lp$ and $\as$, they imply the recursion relations
\begin{align}
F_{i\ib}^{(n+1,l+1)}  = 
\frac{1}{l+1} \bigg[ 
&
\delta_{l, 0} \Gamma^{i}_{n}
+ 
\sum_{s=0}^{n} s\, \beta_{n-s} F_{i\ib}^{(s,l)}
\bigg]
\,\raisebox{-5pt}{,}
\label{eq:recursion_F}
\\
I_{i/j}^{(n+1,l+1)}\!(z)  = 
\frac{1}{l+1} 
\sum_{s=0}^{n}
\bigg[
&
\frac{1}{2}\Gamma^{i}_{n-s} I_{i/j}^{(s,l-1)}\!(z)
+\big(s\, \beta_{n-s} -\gamma^i_{n-s} \big) I_{i/j}^{(s,l)}\!(z)
\nonumber\\
&
-\sum_k I_{i/k}^{(s,l)}\!(z) \otimes P_{k/j}^{(n-s)}(z)
\bigg]
\,\raisebox{-4pt}{.}
\label{eq:recursion_I}
\end{align}
for the coefficients defined above with $n,l\geq 0$. Coefficients on the right hand side with $l'$ outside the range specified in eqs.~(\ref{eq:expand_Fi},\,\ref{eq:expand_Iij}) are understood to vanish.

The coefficients with values $(n',l')$ are thus expressed in terms of coefficients with lower values of $n$ and $l$ and the QCD parameters listed in appendix~\ref{sec:ad}. 
Applying these equations recursively, one can remove all terms with $l>0$ on the right hand sides of these equations as can be shown easily by induction. 
Phrased differently, the functional dependence on $\lp$ of $F_{i\ib}$ and $I_{i/j}$ can be recovered from their values at $\lp=0\,$, $\, F_{i\ib}^{(n,0)}\!=F_{i\ib}^{(n)}(\lp=0)$ and $I_{i/j}^{(n,0)}(z)=I_{i/j}^{(n)}(z,\lp=0)$.

Note that the RGEs also imply $F_{i\ib}^{(0,l)}=0$ as well as $I_{i/j}^{(0,l)}=0$ for $l>0$. From these values and eqs.~(\ref{eq:recursion_F},\,\ref{eq:recursion_I}) the maximal power of $\lp$ per power of $\as$ follows as specified in eqs.~(\ref{eq:expand_Fi},\,\ref{eq:expand_Iij}).

\subsection{Results at lower order}
\label{sec:lower_order}
The anomaly coefficients are obtained as
\begin{align}
& F^{(0)}_{i\ib}(L_\perp) = 0\,,
\hspace{1.8cm}
\frac{F^{(1)}_{q\bar{q}}(0)}{C_F} =
\frac{F^{(1)}_{gg}(0)}{C_A} =
0
\,.
\end{align}
The NNLO results have been given in \eqn{eq:F2_result} and the terms containing $\lp$ are implied by \eqn{eq:recursion_F}. The full results agree with \cite{Becher:2010tm}.

The renormalized matching kernels up to NLO are
\begin{align}
I^{(0)}_{i/j}(z,L_\perp) = &\, \delta_{ij}\delta(1-z)\,,\nonumber\displaybreak[1]\\
I^{(1)}_{g/g}(z,0) = &\,
-C_A \zeta_{2} \delta(1 -z)
\,,\nonumber\displaybreak[1]\\
I^{(1)}_{g/q}(z,0) = &\,
2 C_F z
\,,\nonumber\displaybreak[1]\\
I^{(1)}_{q/g}(z,0) = &\,
2 T_F z ( 2 -z)
\,,\nonumber\displaybreak[1]\\
I^{(1)}_{q/q}(z,0) = &\,
C_F\Big[2 (1 -z) -\zeta_{2} \delta(1 -z) \Big] 
\,,\nonumber\displaybreak[1]\\
I^{(1)}_{q'/q}(z,0)\,,\;& I^{(1)}_{\bar{q}/q}(z,0) = 0
\,.
\label{eq:Ir_1_ren}
\end{align}
Due to eqs.~(\ref{eq:renorm_I},\,\ref{eq:Z_B}) we actually determined them up to $\epsilon^2$ for the extraction of the renormalized NNLO matching kernels.
The terms containing $\lp$ are implied by \eqn{eq:recursion_I}.
The NNLO results have been presented in section~\ref{sec:results}.
The full results are available in the FORM module \cite{Module} associated with this paper.
\subsection{Bare TPDFs}
For the logarithms associated with the analytic regulator we identify 
\begin{align}
 L_a = \log \frac{\nu}{n \tcdot \bar{p}}
 \;\;\;\;\text{ and }\;\;\;
 L_c = \log \frac{\nu\, \nb \tcdot p\,x_T^2}{ 4e^{-2\gamma_E}}
 \label{eq:logs_nu}
\end{align}
for the anti-collinear and collinear region, respectively.

Then the exact NLO results for the bare TPDFs of the collinear and anti-collinear region are given by
\begin{align}
&\Bcal^{(1)}_{i/j}{ (z,x_T^2,\mu,\nu)} 
=
e^{\alpha L_c +\epsilon L_\perp }
e^{- (\epsilon+2\alpha)\gamma_E}
  \frac{\Gamma{ (-\epsilon-\alpha)}}{\Gamma{ (1+\alpha)}} 
({ 1-z})^{\alpha}\, f_{i/j}^{(1)}(z)
\,,\nonumber\\
\label{eq:sB_AC_nlo}
&
\bar{\Bcal}^{(1)}_{i/j}
{ (z,x_T^2,\mu,\nu)} 
= 
e^{\alpha L_a +\epsilon L_\perp}
e^{- \epsilon \gamma_E}
 \Gamma(-\epsilon)
({ 1-z})^{-\alpha} 
\,f_{i/j}^{(1)}(z)
\,,
\end{align}
 with the functions
\begin{align}
&f_{g/g}^{(1)}(z)=4C_A \frac{1}{1-z}
\bigg[\frac{\left(1-z+z^2\right)\!\vphantom{z}^2}{z}\bigg]\,\raisebox{-5pt}{,}
\nonumber\displaybreak[1]\\&
f_{g/q}^{(1)}(z)=2C_F\bigg[\frac{1+(1-z)^2}{z}-\epsilon z\bigg]\,\raisebox{-5pt}{,}
\displaybreak[2]
\nonumber\\&
f_{q/g}^{(1)}(z)=2T_F \bigg[1-\frac{2}{1-\epsilon}z(1-z)\bigg]\,\raisebox{-5pt}{,}
\displaybreak[2]
\nonumber\\&
f_{q/q}^{(1)}(z)=2C_F \frac{1}{1-z}
\Big[1+z^2 -\epsilon (1-z)^2\Big]\,\raisebox{-5pt}{,}
\vphantom{\bigg]}
\nonumber\\&
f_{q'\!/q}^{(1)}(z)\, ,\;f_{\bar{q}/q}^{(1)}(z) = 0\,.
\label{eq:NLO_bare_results_exact}
\end{align}
Note that here the coefficients are given with respect to the renormalized coupling constant, such that appropriate powers of the \msbar\ factor are included and the NNLO expressions contains not only the double real and  virtual-real contributions, but also a counter term contribution:
\begin{align}
 \label{eq:sB2_sum}
 \Bcal^{(2)}_{i/j} = 
 \Bcal^{(2,2)}_{i/j} + \Bcal^{(2,1)}_{i/j} -\frac{\beta_0}{\epsilon}\Bcal^{(1)}_{i/j}\,,
\end{align}
and correspondingly for the anti-collinear functions.
The last term is obtained from the NLO results above. 
For the other terms we only list the pole terms in the analytic regulator, since the full results are very lengthy.
The complete results can be found in the FORM module \cite{Module} accompanying this article.
They can be written as
\begin{align}
&\Bcal^{(2,n_r)}_{i/j}{ (z,x_T^2,\mu,\nu)} 
=
e^{n_r \alpha L_c +2\epsilon L_\perp }
\, f_{i/j}^{(2,n_r)}(z,1)
+\sO(\alpha,\epsilon)
\,,\nonumber\\
\label{eq:sB_AC}
&\bar{\Bcal}^{(2,n_r)}_{i/j}{ (z,x_T^2,\mu,\nu)} 
= 
e^{n_r\alpha L_a +2\epsilon L_\perp}
\,f_{i/j}^{(2,n_r)}(z,-1)
+\sO(\alpha,\epsilon)
\,,
\end{align}
with $n_r$ the number of emitted partons.
For the virtual real contribution we identify
\begin{align}
\nonumber
&f_{g/g}^{(2,1)}(z,s) =
C_A^2
\delta(1-z)
\frac{4s}{\alpha}
\bigg\{
 \frac{1}{\epsilon^3} -\frac{1}{\epsilon} \zeta_{2} +\frac{2}{3} \zeta_{3}
\bigg\}
+\sO(\alpha^0)
\,,
\\
\nonumber
&f_{q/q}^{(2,1)}(z,s) =
C_FCC_A
\delta(1-z)
\frac{4 s}{\alpha}
\bigg\{
\frac{1}{\epsilon^3} -\frac{1}{\epsilon} \zeta_{2} +\frac{2}{3} \zeta_{3}
\bigg\}
+\sO(\alpha^0)
\,,
\displaybreak[2]\\
\nonumber
&f_{g/q}^{(2,1)}(z,s)\,,\;f_{q/g}^{(2,1)}(z,s)  = \sO(\alpha^0)\,,
\\
&f_{\bar{q}/q}^{(2,1)}(z,s) = f_{q'/q}^{(2,1)}(z,s) = 0\,.
\end{align}
For the double real contribution the pole terms are obtained as
\begin{align}
&f_{g/g}^{(2,2)}(z,s) =
C_A^2\bigg\{\delta(1-z)\bigg[
\frac{8}{\alpha^2 \epsilon^2} +\frac{8}{\alpha^2} \zeta_{2} -\frac{8 +10 s}{\alpha \epsilon^3} -\frac{11 s}{3 \alpha \epsilon^2}  +\frac{1}{\alpha \epsilon} \Big( -\frac{67 s}{9} +4 s \zeta_{2} \Big)
\nonumber\\
&\quad
+\frac{1}{\alpha} \Big( -\frac{11 s}{3} \zeta_{2} -\frac{404 s}{27} +\frac{2 (4 +23 s)}{3} \zeta_{3}\Big)
\bigg]
+
\tilde{p}_{gg}(z)\bigg[
\frac{16 s}{\alpha \epsilon^2}
+\frac{16 s}{\alpha} \zeta_{2}
\bigg]
\bigg\}
\nonumber\\
&\quad+C_A T_F N_f
\delta(1-z)\bigg\{
\frac{4 s}{3 \alpha \epsilon^2} +\frac{20 s}{9 \alpha \epsilon} 
+\frac{1}{\alpha} \Big(\frac{4 s}{3} \zeta_{2} +\frac{112 s}{27}\Big)
\bigg\}
+\sO(\alpha^0)
\,,
\nonumber
\displaybreak[2]\\
&f_{g/q}^{(2,2)}(z,s) =
C_F C_A
\bigg\{\tilde{p}_{gq}(z)\bigg[
\frac{8 s}{\alpha \epsilon^2} +\frac{8 s}{\alpha} \zeta_{2}
\bigg]
-\frac{8 s z}{\alpha \epsilon} 
\bigg\}
+\sO(\alpha^0)
\,,
\nonumber
\displaybreak[2]\\
&f_{q/g}^{(2,2)}(z,s) =
C_F T_F
\bigg\{\tilde{p}_{qg}(z)\bigg[
 \frac{8 s}{\alpha \epsilon^2} +\frac{8 s}{\alpha \epsilon} +\frac{8s}{\alpha} \big(1 + \zeta_{2} \big) 
 \bigg]
 +\bigg[
-\frac{8 s}{\alpha \epsilon} -\frac{8 s}{\alpha} 
\bigg]
\bigg\}
+\sO(\alpha^0)
\,,
\displaybreak[2]\nonumber
\\
&f_{q/q}^{(2,2)}(z,s) =
C_F C_A
\delta(1-z)\bigg\{
-\frac{2 s}{\alpha \epsilon^3} -\frac{11 s}{3 \alpha \epsilon^2} +\frac{s}{\alpha \epsilon} \Big( 4 \zeta_{2} -\frac{67}{9} \Big) 
+\frac{s}{\alpha} \Big(  \frac{38 }{3} \zeta_{3} -\frac{11 }{3} \zeta_{2} -\frac{404 }{27}\Big)
\bigg\}
\nonumber\displaybreak[1]\\
&\quad
+C_F^2
\bigg\{\delta(1-z)\bigg[
\frac{8}{\alpha^2 \epsilon^2} +\frac{8}{\alpha^2} \zeta_{2} 
-\frac{8(1+s)}{\alpha \epsilon^3}  +\frac{8 (1 +s)}{3 \alpha} \zeta_{3} 
\bigg]
+\tilde{p}_{qq}(z)\bigg[
\frac{8 s}{\alpha \epsilon^2} +\frac{8 s}{\alpha} \zeta_{2} 
\bigg]
\nonumber\\
&\qquad\quad\quad
-\frac{8 (1 -z) s}{\alpha \epsilon}
\bigg\}
+C_FT_F N_f
\delta(1-z)\bigg\{
\frac{4 s}{3 \alpha \epsilon^2} +\frac{20 s}{9 \alpha \epsilon} +\frac{1}{\alpha} \Big(\frac{4 s}{3} \zeta_{2} +\frac{112 s}{27}\Big) 
\bigg\}
+\sO(\alpha^0)
\,,
\nonumber
\displaybreak[1]\\
&f_{q'/q}^{(2,2)}(z,s)\,,\; f_{\bar{q}/q}^{(2,2)}(z,s) = \sO(\alpha^0)
\,.
\end{align}
Using the results listed above, it is a straightforward exercise to confirm the cancellation of all poles in the analytic regulator up to NNLO in $\as$ on the left hand side of \eqn{eq:refactorization} as well as the cancellation of the related scale $\nu$ and the associated generation of the hard scale $q^2\sim \nb\tcdot p\, n\tcdot\bar{p}$ by the difference of the two logarithms in \eqn{eq:logs_nu}.

\bibliographystyle{JHEP}
\bibliography{paper}

\end{document}